\documentclass{ametsocV6.1_4arxiv}

\usepackage{comment}
\usepackage{url}
\title{Transformer-based nowcasting of radar composites from satellite images for severe weather}

\authors{\c{C}a\u{g}lar~K\"u\c{c}\"uk\aff{a}\correspondingauthor{\c{C}.~K\"u\c{c}\"uk, caglar.kucuk@geosphere.at \\ %\vspace*{2cm}
		{\normalsize \textbf{This article has been accepted to the AMS journal \textit{Artificial Intelligence for the Earth Systems}.}}
	}, Apostolos Giannakos\aff{a}, Stefan Schneider\aff{a}, Alexander Jann\aff{a}} 
\affiliation{\aff{a}{GeoSphere Austria -- Federal Institute for Geology, Geophysics, Climatology and Meteorology, Vienna, Austria}}

\abstract{ % 260 words max!
Weather radar data are critical for nowcasting and an integral component of numerical weather prediction models. While weather radar data provide valuable information at high resolution, their ground-based nature limits their availability, which impedes large-scale applications. In contrast, meteorological satellites cover larger domains but with coarser resolution.
However, with the rapid advancements in data-driven methodologies and modern sensors aboard geostationary satellites, new opportunities are emerging to bridge the gap between ground- and space-based observations, ultimately leading to more skillful weather prediction with high accuracy.
Here, we present a Transformer-based model for nowcasting ground-based radar image sequences using satellite data up to two hours lead time. Trained on a dataset reflecting severe weather conditions, the model predicts radar fields occurring under different weather phenomena and shows robustness against rapidly growing/decaying fields and complex field structures.
Model interpretation reveals that the infrared channel centered at 10.3 $\mu m$ (C13) contains skillful information for all weather conditions, while lightning data have the highest relative feature importance in severe weather conditions, particularly in shorter lead times.
The model can support precipitation nowcasting across large domains without an explicit need for radar towers, enhance numerical weather prediction and hydrological models, and provide radar proxy for data-scarce regions. Moreover, the open-source framework facilitates progress towards operational data-driven nowcasting.
}

\begin{document}

%TC:ignore

%% Necessary!
\maketitle

%%%%%%%%%%%%%%%%%%%%%%%%%%%%%%%%%%%%%%%%%%%%%%%%%%%%%%%%%%%%%%%%%%%%%
% SIGNIFICANCE STATEMENT/CAPSULE SUMMARY
%%%%%%%%%%%%%%%%%%%%%%%%%%%%%%%%%%%%%%%%%%%%%%%%%%%%%%%%%%%%%%%%%%%%%
%
% Significance Statement (all journals except BAMS)
%
\statement
% No more than 120 words! See \url{www.ametsoc.org/index.cfm/ams/publications/author-information/significance-statements/} for details.

Ground-based weather radar data are essential for nowcasting, but data availability limitations hamper usage of radar data across large domains. 
We present a machine learning model, rooted in Transformer architecture, that performs nowcasting of radar data using high-resolution geostationary satellite retrievals, for lead times of up to two hours. 
Our model captures the spatiotemporal dynamics of radar fields from satellite data and offers accurate forecasts. Analysis indicates that the infrared channel centered at \(10.3 \, \mu m\) provides useful information for nowcasting radar fields under various weather conditions. However, lightning activity exhibits the highest forecasting skill for severe weather at short lead times. Our findings show the potential of Transformer-based models for nowcasting severe weather.

%%%%%%%%%%%%%%%%%%%%%%%%%%%%%%%%%%%%%%%%%%%%%%%%%%%%%%%%%%%%%%%%%%%%%
% MAIN BODY OF PAPER
%%%%%%%%%%%%%%%%%%%%%%%%%%%%%%%%%%%%%%%%%%%%%%%%%%%%%%%%%%%%%%%%%%%%%
%

%TC:endignore
\section{Introduction}
%%%% 1) Summarise the task and its relevance!!!
Nowcasting is crucial for weather-dependent decision-making, ranging from everyday occurrences with high socio-economic implications like aviation and outdoor activities, to extreme events requiring disaster management.
Despite the significant capacity built up over the last few decades \citep{Bauer2015}, challenges remain in nowcasting, particularly in resolving convection at early stages \citep{Yano2018}. 
Weather radar data are critical for nowcasting, numerical weather prediction (NWP), and hydrology models solving rainfall-runoff dynamics \citep{Sokol2021}. 
It can also benefit climate and wildfire studies when used effectively \citep{Saltikoff2019,McCarthy2019}. However, the availability of radar data is limited due to multiple factors such as the operational costs of ground-based stations, irregularities in sensor properties, error handling, and data dissemination among sites. This problem becomes particularly acute in studies covering large domains and data-scarce regions.

%%%% 2) Talk about the opportunities!!!
Conversely, satellite-based products are increasingly valuable in Earth system science and offer enhanced capabilities for Earth observation. Sensors with higher spatiotemporal resolution enable solving of complex problems, such as convective activity across large domains. However,  the large data volumes present particular challenges necessitating the development and application of data-driven methods \citep{Bauer2021, Chantry2021}. 

%%%% 3) Elaborate what has been done so far!!!
Nowcasting radar images involves a space-time prediction problem, which is commonly addressed by integrating convolutional neural networks (CNNs) with recurrent neural networks (RNNs) within the conventional neural networks landscape. 
Powerful methods using temporal state captured by RNNs on two-dimensional convolutional operators were developed to tackle this problem such as convolutional long short-term memory \citep[ConvLSTM;][]{Shi2015}, trajectory gated recurrent unit \citep[TrajGRU;][]{Shi2017}, and predictive RNN \citep[PredRNN;][]{PredRNN}. In addition, \citet{Ravuri2021} proposed a deep generative model consisting of one generator with a ResNet-style ConvGRU and two discriminators with two- and three-dimensional CNN architectures for space and time, which yielded accurate predictions with fine-scale spatial structures. \citet{Earthformer} also presented a novel approach to radar nowcasting using a Transformer-based architecture, computing cuboid attentions in small spatiotemporal blocks where spatiotemporal dependencies are learned across scales. Other noteworthy examples modify established architectures to tailor for their specific targets \citep[e.g., ][]{SEVIR,Ayzel2020,Franch2020}.  

%%%% 4) Explain what we've done to close the 'knowladge gap'!!!
Despite its relevance and being an active topic of research, to the best of our knowledge, data-driven weather radar nowcasting without using any radar data during inference remains an open line of research. 
Leveraging state-of-the-art methods to nowcast radar data using satellite-based products would utilize the information within remote sensing products across large domains, nowcast precipitation with improved skill, lead time, and resolution, and provide more input data for better support other models solving NWP or hydrology problems. Therefore, the main objective of this research is to develop a data-driven model for nowcasting radar mosaics using geostationary satellite products up to 2 hours of lead time with high spatiotemporal resolution (1 kilometer in space, 5 minutes in time). We analyze the model performance through individual events and across a test dataset and present further insights from the model by interpreting the model with permutation tests.

\section{Data and methods}
\subsection{Data}\label{ssec:Data}
We use the Storm EVent ImagRy (SEVIR) dataset \citep{SEVIR}, which combines remote sensing retrievals from the Geostationary Environmental Satellite System (GOES) with ground-based radar mosaics from the Next-Generation Radar (NEXRAD) system. 
The satellite-based part of the dataset contains one visible channel, namely C02 (central wavelength of 0.64 $\mu m$ and a spatial resolution of 0.5 km at nadir), two infrared channels, namely C09 and C13 (central wavelengths of 6.9 $\mu m$ and 10.3 $\mu m$, respectively, and a spatial resolution of 2 km), and lightning counts (8 km when mapped into gridded pixels).

The C09 channel, commonly referred to as the `midlevel water vapor' channel, is used for tracking middle tropospheric winds, identifying jet streams.
The C13 channel, also known as the `clean' longwave window, is the least sensitive infrared channel to water vapor among the ABI channels \citep{Schmit2018}.
Therefore, it is widely used in various applications related to cloud and other atmospheric features, such as estimating cloud-top temperature, cloud particle size, and atmospheric moisture corrections. 
Finally, the lightning counts provides the total number of inter-cloud and cloud-to-ground flashes aggregated over 5 minutes.

The vertically integrated liquid (VIL) mosaics provided by the NEXRAD system, available at 1 km spatial resolution, are the ground-based part in the SEVIR repository.
VIL is a critical element in weather prediction and aviation operations \citep{Smith2016}, as it provides an estimate of the total amount of liquid water in a given atmospheric column.
VIL is an important diagnostic tool for severe weather, especially for hail \citep{graham1999vil}, and it is used in operational nowcasting systems, especially for thunderstorm cells \citep{James2018}. To address the highly skewed distribution of VIL values, the values were nonlinearly transformed to a range of [0-255] in the SEVIR repository \citep{SEVIR}. We used the transformed values during training, but used VIL values from their original domain in the analysis to provide meteorologically relevant results.

All variables in the dataset are available at a 5-minutes temporal resolution.
The SEVIR dataset comprises approximately 13000 patches of image sequences distributed across the United States with a temporal span of 2018-2019. Each patch covers a spatial area of 384 km × 384 km at the native resolution of the channels and a temporal duration of 4 hours, consisting of 49 frames. To circumvent the problem of class imbalance during model training due to the prevalence of clear-sky conditions, \citet{SEVIR} employed stratified sampling of random events, constituting approximately 80\% of the dataset, based on VIL values within the domain. 
Furthermore, SEVIR contains storm events that are sampled from an external storm database cataloged by the National Centers for Environmental Information\footnote{\url{https://www.ncdc.noaa.gov/stormevents}} (NCEI), which enable evaluating model performance on specific storm events. 

We refrained from using the visible channel to ensure comparable model performance for day and night conditions, despite its potential to improve model performance. We normalized each channel by subtracting its mean value and dividing it by the standard deviation, both of which were computed over the training data. Finally, to ensure robust results during analysis, we considered samples from the second half of 2019 as the test data and used the remaining for training the model. 
Of the 11728 events remained after filtering them to have no missing data in the four channels used, 27 \% of the events were used for testing while 70 and 3 \% were used for training and validation, respectively.

\subsection{Methods}\label{ssec:Methods}
We employed a modified version of the Earthformer algorithm \citep{Earthformer}, which is based on the Transformer architecture. Transformers were originally developed for machine translation tasks \citep{Vaswani2017}, and they have proven to be highly effective in time-dependent prediction tasks thanks to their attention mechanism, which can model temporal dependencies across different scales. Transformer-based models have also been applied to two-dimensional computer vision tasks by leveraging the flexibility of the attention mechanism. For instance, \citet{Dosovitskiy2021} demonstrated that Transformers can perform image classification tasks that were traditionally done using CNN-based methods by computing attention on flattened images, which are sequences of image patches.

Moreover, \citet{Earthformer} introduced an efficient patching framework for three-dimensional data by computing attention on small cuboids to perform space-time prediction tasks from multiple domains, such as video prediction and nowcasting. 
The Earthformer architecture utilizes multiple self- and cross-attention blocks to attend the cuboids patched with different strategies, which are connected to each other with a ResNet-style architecture to utilize information available across a hierarchical encoder-decoder structure. The model learns statistical relationships within and between cuboids with self- and cross-attention, respectively. Earthformer has been found to outperform popular space-time prediction models in nowcasting radar image sequences given past data \citep{Earthformer}. Recent studies have also demonstrated that Earthformer is the state-of-the-art in Transformer-based spatiotemporal prediction tasks in the Earth system science \citep[e.g.,][]{Li2023_w4c, Benson2023}.

The objective of the model is to predict future time steps of VIL image sequences spanning over 120 minutes (24 frames) at a spatial resolution of 1km, given the past satellite image sequences of two infrared channels and lightning counts spanning over 125 minutes (25 frames). To match the resolution of the input data with the target variable in the data pipeline, we increased the spatial resolution of the input data to 1km using nearest neighbor interpolation on the infrared channels (2km) and lightning counts (8km). Moreover, the Earthformer architecture was modified through a limited exploratory search for hyperparameters, guided by domain expertise. To enhance the model's capacity to learn complex interactions between satellite and radar data, we increased the number of attention heads to 16, where each head is expected to perform self-attention on different scales of motion. Similarly, we increased the number of global vectors to 32 to ensure that the cross-attention between different cuboids was incorporated into the model. After modifications, the model had 12.5 million trainable parameters.

We used the original loss function, mean squared error, proposed by \citet{Earthformer}, but scaled the loss values across time dimension with an inverse linear function to give greater weight to earlier time steps of the model output. 
We used a warmed-up learning rate that increases to 5x10$^{-4}$ during the first 15 \% of the maximum number of epochs, which is set to 120, then decreases with cosine annealing. We used the validation set to trigger early stopping of the model training, which stopped training in 106th epoch. One epoch took around 1 hour on a virtual machine with a GPU of 48 GB memory on an RTX A6000 graphics card. 
%a virtual GPU of 48 GB RAM on an RTX A6000. 
To compare the model with a simple baseline, we trained a 3D U-Net \citep{Cicek2016} with comparable model size (14.8 million trainable parameters) using the same optimization architecture. Further details are available in the repository accompanying the article.

\section{Results}

\subsection{Model performance}

%TC:ignore
\begin{figure*}[!h] % Model output
	\includegraphics[width=\linewidth,keepaspectratio]{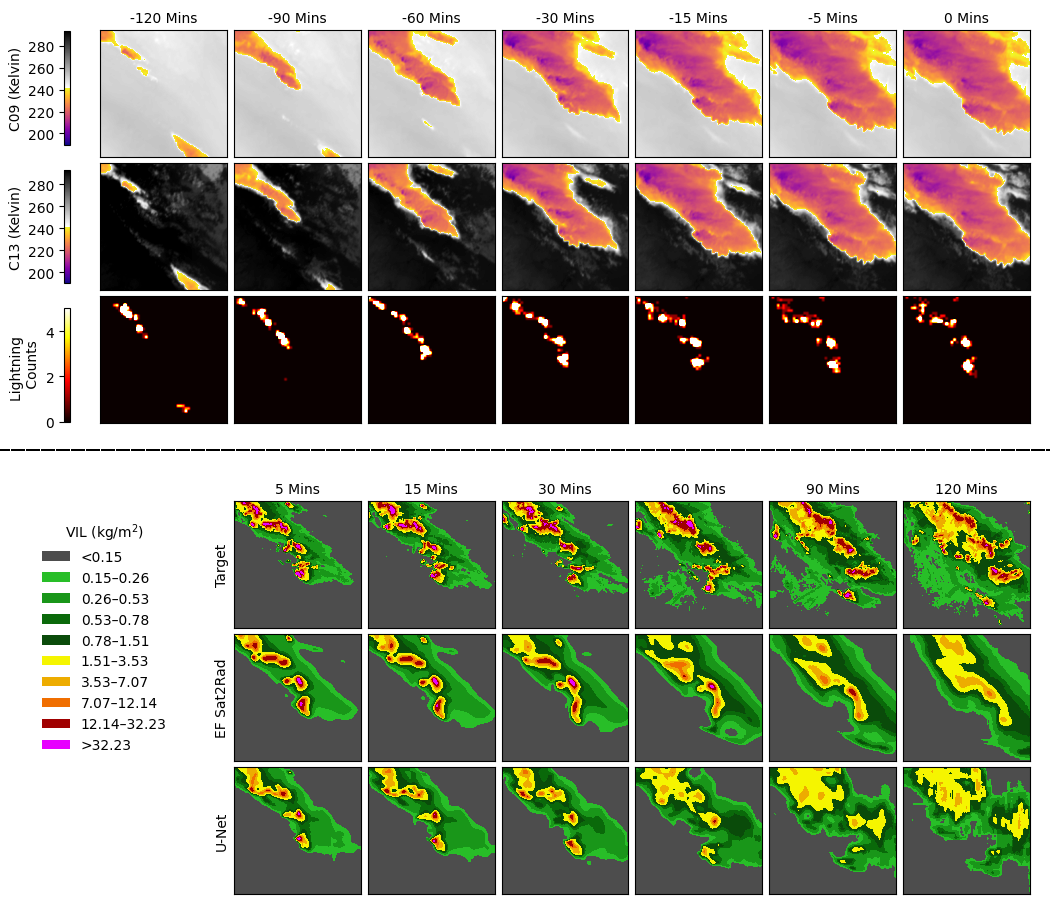}
	
	\caption{A thunderstorm wind event from test dataset (ID=S857496 the in SEVIR catalog, also accessible through the NCEI catalog via \protect\url{https://www.ncdc.noaa.gov/stormevents/eventdetails.jsp?id=857496}). Input data are shown in the first three rows for the channels C09 (6.9 $\mu m$), C13 (10.3 $\mu m$), and lightning counts, respectively. Target VIL fields are shown in the fourth row, while model predictions for our model (EF Sat2Rad) and the baseline model (U-Net) are shown in the fifth and sixth rows, respectively. Time steps are given at the top of both input and output data, separated with a dashed line. Note that temporal intervals for input and output are 5 minutes, but plotted with varying intervals for visualization purposes.}
	\label{fig:modelOut}
\end{figure*}
%TC:endignore

Model performance for a single event is illustrated in Fig. \ref{fig:modelOut}, where the input data are presented in the first three rows, the target in the fourth, and the predictions of our modified Earthformer model (EF Sat2Rad) and the 3D U-Net model as the baseline in the last two rows, respectively. Temperature variations in space are evident in both infrared channels of the input data, albeit sharper in C13, and it overlaps with lightning activity in the early stages of the convection.  Regarding the target VIL values, multiple storm cells with large VIL values have developed within the earlier lead times and later moved eastward. Our model successfully predicted the structure and spatial extent of the storm cells, showed superior performance than the baseline. Despite missing the scale of the largest VIL values, our model captured the locations with high activity, especially in the earlier lead times. While both models show a tendency to smooth predicted fields with increasing lead time, performance of our model degrades slower than the baseline, indicating a better capture of temporal dependencies. 
Overall, given the problem's complexity, the high spatial resolution, and the extended lead time, our model successfully learned the statistical relationships between satellite retrievals and ground-based VIL mosaics. Moreover, our model effectively reproduced the fields' growth, decay, and temporal movement (see supplements for additional examples).

\begin{figure*}%[b] % Model performance quantified
	\centering
	\includegraphics[width=.6\linewidth,keepaspectratio]{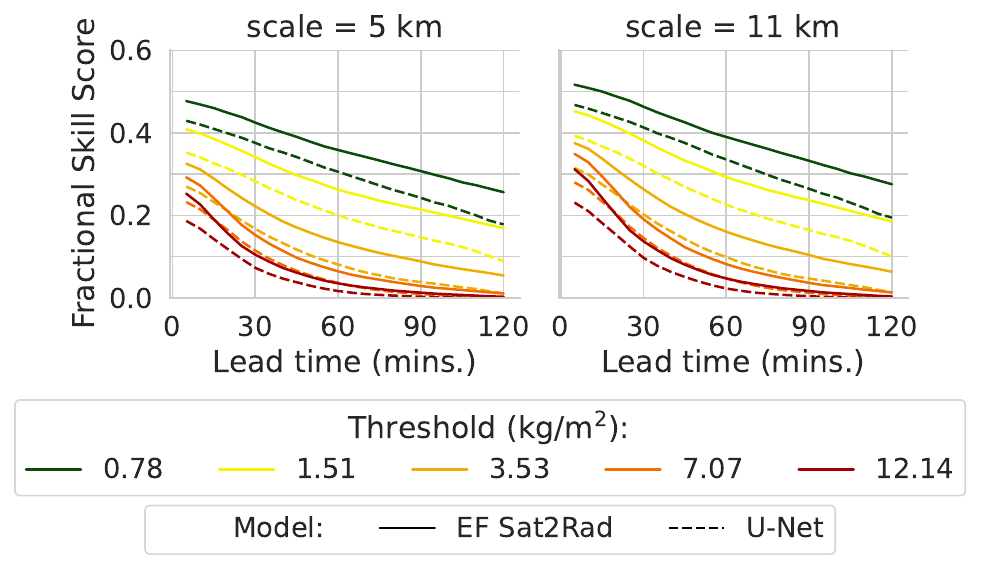}
	\caption{Summary of model performance using FSS at different spatial scales for our model and the baseline, indicated by solid and dashed lines, respectively.}
	\label{fig:modelPerf}
\end{figure*}

We quantified model performance over the entire test set using Fractional Skill Score \citep[FSS;][]{Roberts2008} on full image patches with spatial scales of 5 and 11 km to consider neighboring pixels, with several thresholds ranging from 0.78 to 12.14 kg/m$^2$ of VIL, parallel to the ones used in \citet{SEVIR}. Mean FSS values per time step are shown in Fig. \ref{fig:modelPerf} for our model and the baseline in solid and dashed lines, respectively. Our model outperforms the baseline in all cases. Decrease in performance with increasing lead time is observed in both models, though to a lesser extent in our model.

%TC:ignore
%TC:endignore

\subsection{Model interpretation}

To gain confidence in model predictions and diagnose importance of channels and temporal extent of the input data, we conducted two permutation tests by shuffling input of the test dataset across channel and time. 
%We attributed the skill score, normalized reduction in FSS values computed with 5km spatial scale (summarized in Fig. \ref{fig:modelPerf}), to importance of the permuted chunk. 
We computed the skill score as $1 - FSS_{X}/FSS_{ref}$ where $FSS_{ref}$ is the score of the original test set with 5km spatial scale (summarized in Fig. \ref{fig:modelPerf}) and $FSS_{X}$ is the score of the same dataset with the $X$ chunk permuted, and attributed that to importance of the permuted chunk. 

%TC:ignore
\begin{figure*}%[b] % Model interpretation 
	\includegraphics[width=\linewidth,keepaspectratio]{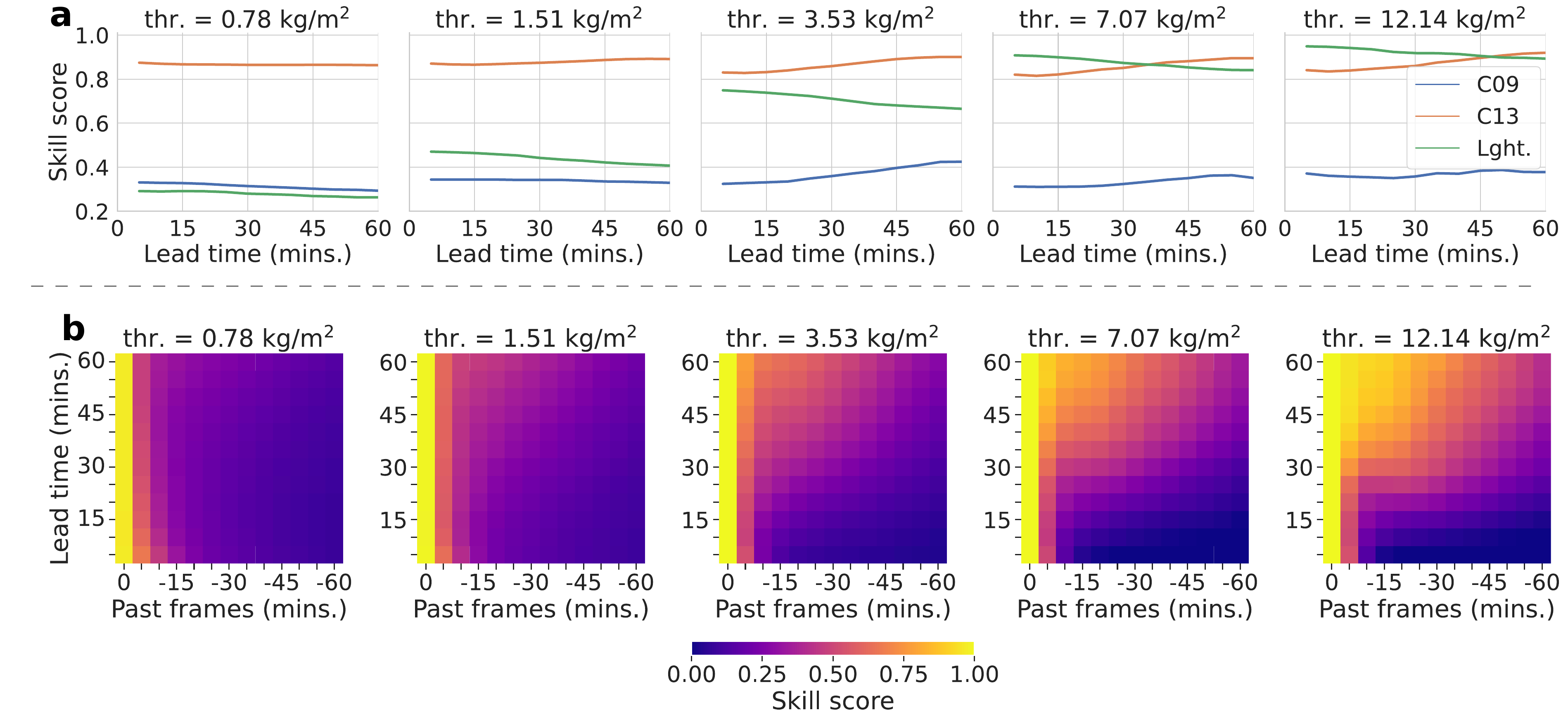}
	\caption{Relative importance of channel and length of input data for different VIL thresholds, quantified via skill score of each permuted chunk compared to the original model performance as $1 - FSS_{X}/FSS_{ref}$ where $FSS_{ref}$ is the score of the original test set (summarized in Fig. \ref{fig:modelPerf}) and $FSS_{X}$ is the score of the same dataset with the $X$ chunk permuted: (a) For input channels (b) For temporal extent of input data.} 
	\label{fig:modelInt}
\end{figure*}
%TC:endignore

Fig. \ref{fig:modelInt}a summarizes feature importance across channels, thresholds, and up to 1-hour lead time. In general, C13 exhibits the highest skill score, particularly in low VIL thresholds. This aligns with C13's known utility in severe weather diagnoses such as analyzing tropical cyclones \citep{Dvorak1984}, detecting overshooting tops \citep{Dworak2012}, and to translate GOES observations to radar reflectivity fields \citep{Hilburn2020}. Overall, the consistently high relative importance values confirm the potential of the C13 channel to sense steep temperature gradients and low temperature values in the upper troposphere, making it a robust tool for detecting unstable weather activity. 

While C13 has the highest skill score in low VIL thresholds, lightning counts have the highest skill score in large VIL thresholds, as indicated in Fig. \ref{fig:modelInt}a.
This finding aligns with prior studies emphasizing the potential of lightning activity in diagnosing severe weather \citep{Deierling2008,Barthe2010,Avila2010}. The slight decrease in the skill score of lightning counts with increasing lead time is consistent with similar studies \citep[e.g.,][]{Leinonen2023}, and suggests that it is most informative in shorter lead times. 
The low skill score of lightning counts in low VIL thresholds likely results from the absence of lightning activity in such conditions. This effectively prevents the model from extracting useful information from the lightning channel in weather scenarios not yielding high VIL values. 
This aligns with the interpretation of a CNN-based model trained to translate GOES-R observations to radar fields, which suggested that the model primarily focuses on regions with lightning activity when available, or with sharp gradients and extremely low values of infrared channels when lightning is not observed \citep{Hilburn2020}.

The skill score of the C09 channel remains the lowest for the majority of VIL thresholds. This is unsurprising since C09 senses the mean temperature value of a given water column \citep{Schmit2018} while the temperature profile within the column can vary significantly.

Importance of the temporal extent of the input data is summarized in Fig. \ref{fig:modelInt}b. Here, skill scores are calculated based on recursively permuted input data across the time dimension, i.e., permuting a range of frames from the oldest (t=-120) to the length of input data to be analyzed.
%%%%%%%%%%%%%%%%%%%%%%%%%%% 
For lower VIL thresholds, only the last frames closest to the prediction time exhibit significant skill for all lead times. For instance, input data at t=0 provides the most predictive power for all lead times, a result that is expected in scenarios without convective activity. Conversely, for higher VIL thresholds, the skill associated with the temporal extent of the input data shows a consistent increase with longer lead times. Input data at t=-45, for example, exhibits predictive skill for t=45 when dealing with larger VIL thresholds. This suggests that the model is capable of capturing early indicators of convection, which are useful for predicting large VIL values even at extended lead times.

\section{Summary}

In this study, we presented a Transformer-based model for nowcasting radar image sequences by using only remote sensing products with up to two hours of lead time and high resolution. 
Leveraging the strength of attention mechanisms in learning statistical relationships across scales by efficiently utilizing in space-time cuboids with a modular architecture, Earthformer provides a strong Transformers-based baseline for addressing Earth system-related problems that contains processes with largely varying spatio-temporal scales. 
Trained on a dataset reflecting severe weather conditions, our model predicted radar fields occurring under various weather phenomena. The model demonstrated robustness against challenges such as rapidly growing/decaying fields and complex field structures. However, the model showed limited performance against smoothing out the predicted fields in longer lead times, a common problem for non-generative data-driven nowcasting models.

We gained insights from the model by performing permutation tests to infer importance of channel and length of the input data. We found that the C13 channel contains skillful information for all weather conditions, while the lightning count channel has the highest relative feature importance in severe weather conditions, particularly in shorter lead times. Furthermore, we found that the model's predictive skill is sensitive to input data length in nowcasting severe weather conditions. While permutation tests can introduce bias and should be interpreted cautiously, our findings corroborate the significance of both the C13 channel and lightning activity in severe weather nowcasting, thereby strengthening confidence in our analysis.

Our Transformer-based model serves as a proof-of-concept, demonstrating the potential for satellite-based precipitation nowcasting over large domains without the necessity of explicit radar tower support, once it is trained. Moreover, model output can be used as a radar proxy in regions with limited data availability. Improving model performance with better downscaling of the input data and expanding the model with more input data remains to be addressed in future studies towards operational data-driven nowcasting.

\clearpage
%%%%%%%%%%%%%%%%%%%%%%%%%%%%%%%%%%%%%%%%%%%%%%%%%%%%%%%%%%%%%%%%%%%%%
% ACKNOWLEDGMENTS
%%%%%%%%%%%%%%%%%%%%%%%%%%%%%%%%%%%%%%%%%%%%%%%%%%%%%%%%%%%%%%%%%%%%%
\acknowledgments
\c{C}K is supported by the European Organisation for the Exploitation of Meteorological Satellites (EUMETSAT) fellowship ``FUSEDCAST: A fused data approach to the nowcasting of severe weather'', hosted by the Federal Institute for Geology, Geophysics, Climatology and Meteorology. We acknowledge the computing resources provided by the European Weather Cloud. We would also like to thank Aitor Atencia, Markus Dabernig, and Irene Schicker for their feedback on earlier versions of the study.

%TC:ignore

%%%%%%%%%%%%%%%%%%%%%%%%%%%%%%%%%%%%%%%%%%%%%%%%%%%%%%%%%%%%%%%%%%%%%
% DATA AVAILABILITY STATEMENT
%%%%%%%%%%%%%%%%%%%%%%%%%%%%%%%%%%%%%%%%%%%%%%%%%%%%%%%%%%%%%%%%%%%%%
% 
%
\datastatement %\label{sec:Repos}
Data used in the experiments are available in repositories given in \citet{SEVIR}. 
Original version of the Earthformer package can be found in the repository given in \citet{Earthformer}, while modified version used in this study is in \url{https://github.com/caglarkucuk/earthformer-satellite-to-radar}. Weights of the trained model can be found in \url{https://zenodo.org/doi/10.5281/zenodo.10033640}.
%%%%%%%%%%%%%%%%%%%%%%%%%%%%%%%%%%%%%%%%%%%%%%%%%%%%%%%%%%%%%%%%%%%%%
% APPENDIXES
%%%%%%%%%%%%%%%%%%%%%%%%%%%%%%%%%%%%%%%%%%%%%%%%%%%%%%%%%%%%%%%%%%%%%
%
%% If only one appendix, use

%%%%%%%%%%%%%%%%%%%%%%%%%%%%%%%%%%%%%%%%%%%%%%%%%%%%%%%%%%%%%%%%%%%%%
% REFERENCES
%%%%%%%%%%%%%%%%%%%%%%%%%%%%%%%%%%%%%%%%%%%%%%%%%%%%%%%%%%%%%%%%%%%%%
% Make your BibTeX bibliography by using these commands:
\clearpage
\bibliographystyle{ametsocV6}
\bibliography{references}

\begin{thebibliography}{32}
\providecommand{\natexlab}[1]{#1}
\providecommand{\url}[1]{\texttt{#1}}
\renewcommand{\UrlFont}{\rmfamily}
\providecommand{\urlprefix}{URL }
\expandafter\ifx\csname urlstyle\endcsname\relax
  \providecommand{\doi}[1]{https://doi.org/\discretionary{}{}{}#1}\else
  \providecommand{\doi}{https://doi.org/\discretionary{}{}{}\begingroup
  \urlstyle{rm}\Url}\fi
\providecommand{\eprint}[2][]{\url{#2}}

\bibitem[{\'Avila et~al.(2010)\'Avila, B\"urgesser, Castellano, Collier,
  Compagnucci,, and Hughes}]{Avila2010}
\'Avila, E.~E., R.~E. B\"urgesser, N.~E. Castellano, A.~B. Collier, R.~H.
  Compagnucci, and A.~R. Hughes, 2010: Correlations between deep convection and
  lightning activity on a global scale. \textit{Journal of Atmospheric and
  Solar-Terrestrial Physics}, \textbf{72~(14)}, 1114--1121,
  \doi{j.jastp.2010.07.019}.

\bibitem[{Ayzel et~al.(2020)Ayzel, Scheffer,, and Heistermann}]{Ayzel2020}
Ayzel, G., T.~Scheffer, and M.~Heistermann, 2020: {RainNet v1.0: a
  convolutional neural network for radar-based precipitation nowcasting}.
  \textit{Geoscientific Model Development}, \textbf{13~(6)}, 1--20,
  \doi{10.5194/gmd-13-2631-2020}.

\bibitem[{Barthe et~al.(2010)Barthe, Deierling,, and Barth}]{Barthe2010}
Barthe, C., W.~Deierling, and M.~C. Barth, 2010: Estimation of total lightning
  from various storm parameters: A cloud-resolving model study. \textit{Journal
  of Geophysical Research: Atmospheres}, \textbf{115~(D24)},
  \doi{10.1029/2010JD014405}.

\bibitem[{Bauer et~al.(2021)Bauer, Dueben, Hoefler, Quintino, Schulthess,, and
  Wedi}]{Bauer2021}
Bauer, P., P.~D. Dueben, T.~Hoefler, T.~Quintino, T.~C. Schulthess, and N.~P.
  Wedi, 2021: {The digital revolution of Earth-system science}. \textit{Nature
  Computational Science}, \textbf{1~(2)}, 104--113,
  \doi{10.1038/s43588-021-00023-0}.

\bibitem[{Bauer et~al.(2015)Bauer, Thorpe,, and Brunet}]{Bauer2015}
Bauer, P., A.~Thorpe, and G.~Brunet, 2015: {The quiet revolution of numerical
  weather prediction}. \textit{Nature}, \textbf{525~(7567)}, 47--55,
  \doi{10.1038/nature14956}.

\bibitem[{Benson et~al.(2023)}]{Benson2023}
Benson, V., and Coauthors, 2023: {Forecasting localized weather impacts on
  vegetation as seen from space with meteo-guided video prediction}.
  \doi{10.48550/arXiv.2303.16198}.

\bibitem[{Chantry et~al.(2021)Chantry, Christensen, Dueben,, and
  Palmer}]{Chantry2021}
Chantry, M., H.~Christensen, P.~Dueben, and T.~Palmer, 2021: {Opportunities and
  challenges for machine learning in weather and climate modelling: Hard,
  medium and soft AI}. \textit{Philosophical Transactions of the Royal Society
  A}, \textbf{379~(2194)}, \doi{10.1098/rsta.2020.0083}.

\bibitem[{{\c{C}}i{\c{c}}ek et~al.(2016){\c{C}}i{\c{c}}ek, Abdulkadir,
  Lienkamp, Brox,, and Ronneberger}]{Cicek2016}
{\c{C}}i{\c{c}}ek, {\"O}., A.~Abdulkadir, S.~S. Lienkamp, T.~Brox, and
  O.~Ronneberger, 2016: {3D U-Net: learning dense volumetric segmentation from
  sparse annotation}. \textit{Medical Image Computing and Computer-Assisted
  Intervention -- MICCAI 2016}, Springer, 424--432,
  \doi{10.1007/978-3-319-46723-8_49}.

\bibitem[{Deierling and Petersen(2008)Deierling, and Petersen}]{Deierling2008}
Deierling, W., and W.~A. Petersen, 2008: {Total lightning activity as an
  indicator of updraft characteristics}. \textit{Journal of Geophysical
  Research Atmospheres}, \textbf{113~(16)}, \doi{10.1029/2007JD009598}.

\bibitem[{Dosovitskiy et~al.(2021)}]{Dosovitskiy2021}
Dosovitskiy, A., and Coauthors, 2021: An image is worth 16x16 words:
  Transformers for image recognition at scale. \doi{10.48550/arXiv.2010.11929}.

\bibitem[{Dvorak(1984)}]{Dvorak1984}
Dvorak, V.~F., 1984: Tropical cyclone intensity analysis using satellite data.
  \textit{NOAA Tech. Rep.}, \textbf{11}, 47,
  \urlprefix\url{https://severeweather.wmo.int/TCFW/RAI_Training/Dvorak_1984.pdf}.

\bibitem[{Dworak et~al.(2012)Dworak, Bedka, Brunner,, and Feltz}]{Dworak2012}
Dworak, R., K.~Bedka, J.~Brunner, and W.~Feltz, 2012: {Comparison between
  GOES-12 overshooting-top detections, WSR-88D radar reflectivity, and severe
  storm reports}. \textit{Weather and Forecasting}, \textbf{27~(3)}, 684--699,
  \doi{10.1175/WAF-D-11-00070.1}.

\bibitem[{Franch et~al.(2020)Franch, Nerini, Pendesini, Coviello, Jurman,, and
  Furlanello}]{Franch2020}
Franch, G., D.~Nerini, M.~Pendesini, L.~Coviello, G.~Jurman, and C.~Furlanello,
  2020: {Precipitation nowcasting with orographic enhanced stacked
  generalization: Improving deep learning predictions on extreme events}.
  \textit{Atmosphere}, \textbf{11~(3)}, \doi{10.3390/atmos11030267}.

\bibitem[{Gao et~al.(2022)Gao, Shi, Wang, Zhu, Wang, Li,, and
  Yeung}]{Earthformer}
Gao, Z., X.~Shi, H.~Wang, Y.~Zhu, Y.~Wang, M.~Li, and D.-Y. Yeung, 2022:
  Earthformer: Exploring space-time transformers for earth system forecasting.
  \textit{Advances in Neural Information Processing Systems}, Vol.~35,
  \urlprefix\url{https://proceedings.neurips.cc/paper_files/paper/2022/file/a2affd71d15e8fedffe18d0219f4837a-Paper-Conference.pdf}.

\bibitem[{Graham and Struthwolf(1999)Graham, and Struthwolf}]{graham1999vil}
Graham, R.~A., and M.~Struthwolf, 1999: {VIL density as a potential hail
  indicator across northern Utah}. \textit{NOAA Western Region Tech.
  Attachment}, 99--02.

\bibitem[{Hilburn et~al.(2020)Hilburn, Ebert-Uphoff,, and Miller}]{Hilburn2020}
Hilburn, K.~A., I.~Ebert-Uphoff, and S.~D. Miller, 2020: {Development and
  interpretation of a neural-network-based synthetic radar reflectivity
  estimator using GOES-R satellite observations}. \textit{Journal of Applied
  Meteorology and Climatology}, \textbf{60~(1)}, 3--21,
  \doi{10.1175/JAMC-D-20-0084.1}.

\bibitem[{James et~al.(2018)James, Reichert,, and Heizenreder}]{James2018}
James, P.~M., B.~K. Reichert, and D.~Heizenreder, 2018: {NowCastMIX: Automatic
  integrated warnings for severe convection on nowcasting time scales at the
  German weather service}. \textit{Weather and Forecasting}, \textbf{33~(5)},
  1413--1433, \doi{10.1175/WAF-D-18-0038.1}.

\bibitem[{Leinonen et~al.(2023)Leinonen, Hamann, Sideris,, and
  Germann}]{Leinonen2023}
Leinonen, J., U.~Hamann, I.~V. Sideris, and U.~Germann, 2023: {Thunderstorm
  nowcasting with deep learning: a multi-hazard data fusion model}.
  \textit{Geophysical Research Letters}, \textbf{50},
  \doi{10.1029/2022GL101626}.

\bibitem[{Li et~al.(2022)Li, Dong, Fang, Weyn,, and Luferenko}]{Li2023_w4c}
Li, Y., H.~Dong, Z.~Fang, J.~Weyn, and P.~Luferenko, 2022: {Super-resolution
  Probabilistic Rain Prediction from Satellite Data Using 3D U-Nets and
  EarthFormers}. \doi{10.48550/arXiv.2212.02998}.

\bibitem[{McCarthy et~al.(2019)McCarthy, Guyot, Dowdy,, and
  McGowan}]{McCarthy2019}
McCarthy, N., A.~Guyot, A.~Dowdy, and H.~McGowan, 2019: {Wildfire and Weather
  Radar: A Review}. \textit{Journal of Geophysical Research: Atmospheres},
  \textbf{124~(1)}, 266--286, \doi{10.1029/2018JD029285}.

\bibitem[{Ravuri et~al.(2021)}]{Ravuri2021}
Ravuri, S., and Coauthors, 2021: {Skilful Precipitation Nowcasting using Deep
  Generative Models of Radar}. \textit{Nature}, \textbf{597},
  \doi{10.1038/s41586-021-03854-z}.

\bibitem[{Roberts and Lean(2008)Roberts, and Lean}]{Roberts2008}
Roberts, N.~M., and H.~W. Lean, 2008: {Scale-selective verification of rainfall
  accumulations from high-resolution forecasts of convective events}.
  \textit{Monthly Weather Review}, \textbf{136~(1)}, 78--97,
  \doi{10.1175/2007MWR2123.1}.

\bibitem[{Saltikoff et~al.(2019)}]{Saltikoff2019}
Saltikoff, E., and Coauthors, 2019: {An overview of using weather radar for
  climatological studies successes, challenges, and potential}.
  \textit{Bulletin of the American Meteorological Society}, \textbf{100~(9)},
  1739--1751, \doi{10.1175/BAMS-D-18-0166.1}.

\bibitem[{Schmit et~al.(2018)Schmit, Lindstrom, Gerth,, and
  Gunshor}]{Schmit2018}
Schmit, T.~J., S.~S. Lindstrom, J.~J. Gerth, and M.~M. Gunshor, 2018:
  {Applications of the 16 spectral bands on the Advanced Baseline Imager
  (ABI)}. \textit{Journal of Operational Meteorology}, \textbf{6},
  \doi{10.15191/nwajom.2018.0604}.

\bibitem[{Shi et~al.(2015)Shi, Chen, Wang, Yeung, Wong,, and Woo}]{Shi2015}
Shi, X., Z.~Chen, H.~Wang, D.-Y. Yeung, W.-k. Wong, and W.-c. Woo, 2015:
  {Convolutional LSTM network: A machine learning approach for precipitation
  nowcasting}. \textit{Advances in Neural Information Processing Systems},
  Vol.~28,
  \urlprefix\url{https://proceedings.neurips.cc/paper_files/paper/2015/file/07563a3fe3bbe7e3ba84431ad9d055af-Paper.pdf}.

\bibitem[{Shi et~al.(2017)Shi, Gao, Lausen, Wang, Yeung, Wong,, and
  Woo}]{Shi2017}
Shi, X., Z.~Gao, L.~Lausen, H.~Wang, D.-Y. Yeung, W.~K. Wong, and W.~C. Woo,
  2017: {Deep learning for precipitation nowcasting: A benchmark and a new
  model}. \textit{Advances in Neural Information Processing Systems}, Vol.~30,
  \urlprefix\url{https://proceedings.neurips.cc/paper_files/paper/2017/file/a6db4ed04f1621a119799fd3d7545d3d-Paper.pdf}.

\bibitem[{Smith et~al.(2016)}]{Smith2016}
Smith, T.~M., and Coauthors, 2016: {Multi-Radar Multi-Sensor (MRMS) severe
  weather and aviation products: Initial Operating Capabilities}.
  \textit{Bulletin of the American Meteorological Society}, \textbf{97~(9)},
  1617--1630, \doi{10.1175/BAMS-D-14-00173.1}.

\bibitem[{Sokol et~al.(2021)Sokol, Szturc, Orellana-Alvear, Popov{\'{a}},
  Jurczyk,, and C{\'{e}}lleri}]{Sokol2021}
Sokol, Z., J.~Szturc, J.~Orellana-Alvear, J.~Popov{\'{a}}, A.~Jurczyk, and
  R.~C{\'{e}}lleri, 2021: {The role of weather radar in rainfall estimation and
  its application in meteorological and hydrological modelling —A review}.
  \textit{Remote Sensing}, \textbf{13~(3)}, 1--38, \doi{10.3390/rs13030351}.

\bibitem[{Vaswani et~al.(2017)Vaswani, Shazeer, Parmar, Uszkoreit, Jones,
  Gomez, Kaiser,, and Polosukhin}]{Vaswani2017}
Vaswani, A., N.~Shazeer, N.~Parmar, J.~Uszkoreit, L.~Jones, A.~N. Gomez,
  {\L}.~Kaiser, and I.~Polosukhin, 2017: Attention is all you need.
  \textit{Advances in Neural Information Processing Systems}, Vol.~30,
  \urlprefix\url{https://proceedings.neurips.cc/paper_files/paper/2017/file/3f5ee243547dee91fbd053c1c4a845aa-Paper.pdf}.

\bibitem[{Veillette et~al.(2020)Veillette, Samsi,, and Mattioli}]{SEVIR}
Veillette, M., S.~Samsi, and C.~J. Mattioli, 2020: {SEVIR : A Storm Event
  Imagery Dataset for Deep Learning Applications in Radar and Satellite
  Meteorology}. \textit{Advances in Neural Information Processing Systems},
  Vol.~33,
  \urlprefix\url{https://proceedings.neurips.cc/paper/2020/file/fa78a16157fed00d7a80515818432169-Paper.pdf}.

\bibitem[{Wang et~al.(2017)Wang, Long, Wang, Gao,, and Yu}]{PredRNN}
Wang, Y., M.~Long, J.~Wang, Z.~Gao, and P.~S. Yu, 2017: {PredRNN: Recurrent
  Neural Networks for Predictive Learning using Spatiotemporal LSTMs}.
  \textit{Advances in Neural Information Processing Systems}, Vol.~30,
  \urlprefix\url{https://proceedings.neurips.cc/paper_files/paper/2017/file/e5f6ad6ce374177eef023bf5d0c018b6-Paper.pdf}.

\bibitem[{Yano et~al.(2018)}]{Yano2018}
Yano, J. I.~I., and Coauthors, 2018: {Scientific challenges of convective-scale
  numerical weather prediction}. \textit{Bulletin of the American
  Meteorological Society}, \textbf{99~(4)}, 699--710,
  \doi{10.1175/BAMS-D-17-0125.1}.

\end{thebibliography}

%TC:endignore
\clearpage
\appendix
\label{sec:appx}

\appendixtitle{Model output examples}

Further examples from the test dataset are given here as selected events for 5 best and 5 worst performance, as well as 10 randomly selected events.

\subsection{Best examples}

%\textbf{Best}

\begin{figure*}[h] % Model output
	\centering
	\includegraphics[width=.7\linewidth,keepaspectratio]{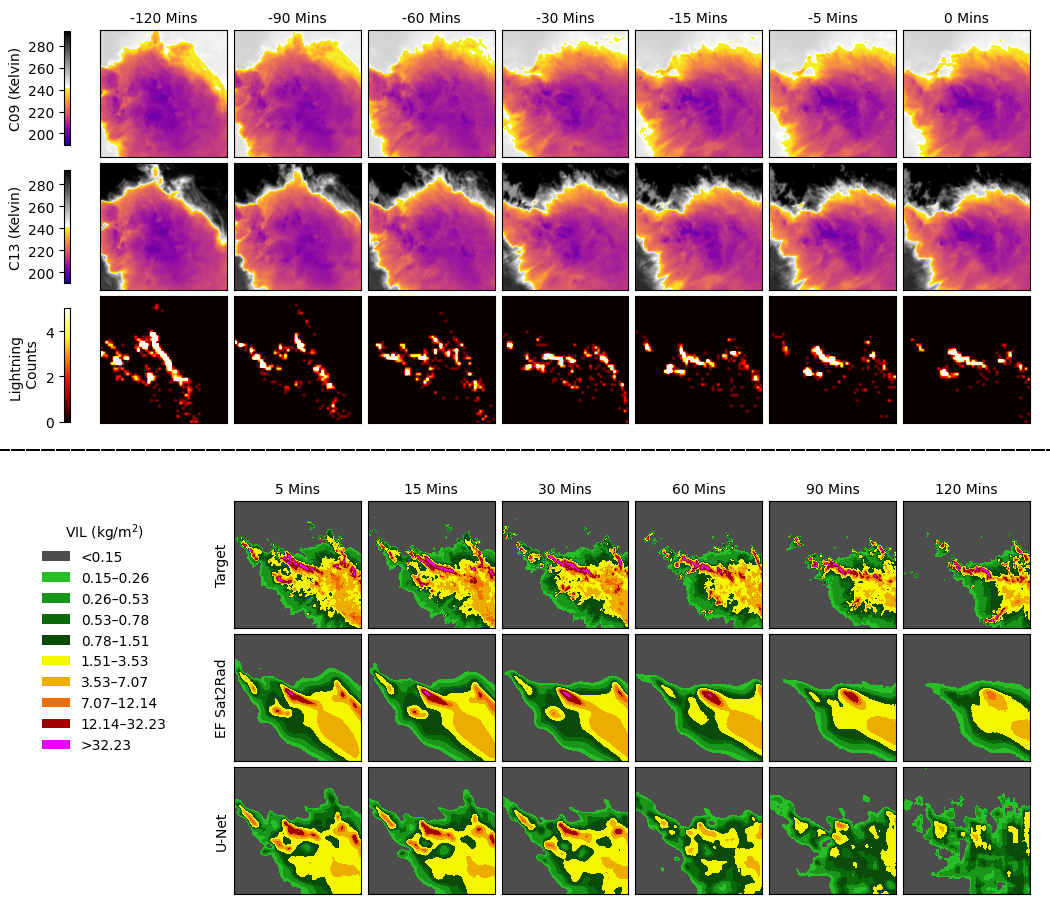}
	\caption{Same as Fig. \ref{fig:modelOut}, but for a hail event with ID=S857225 in the SEVIR catalog.}	% , plotted with the same colorbar,
	%	\label{fig:modelOut_supp1}
\end{figure*}

\begin{figure*}[h] % Model output
	\centering
	\includegraphics[width=.7\linewidth,keepaspectratio]{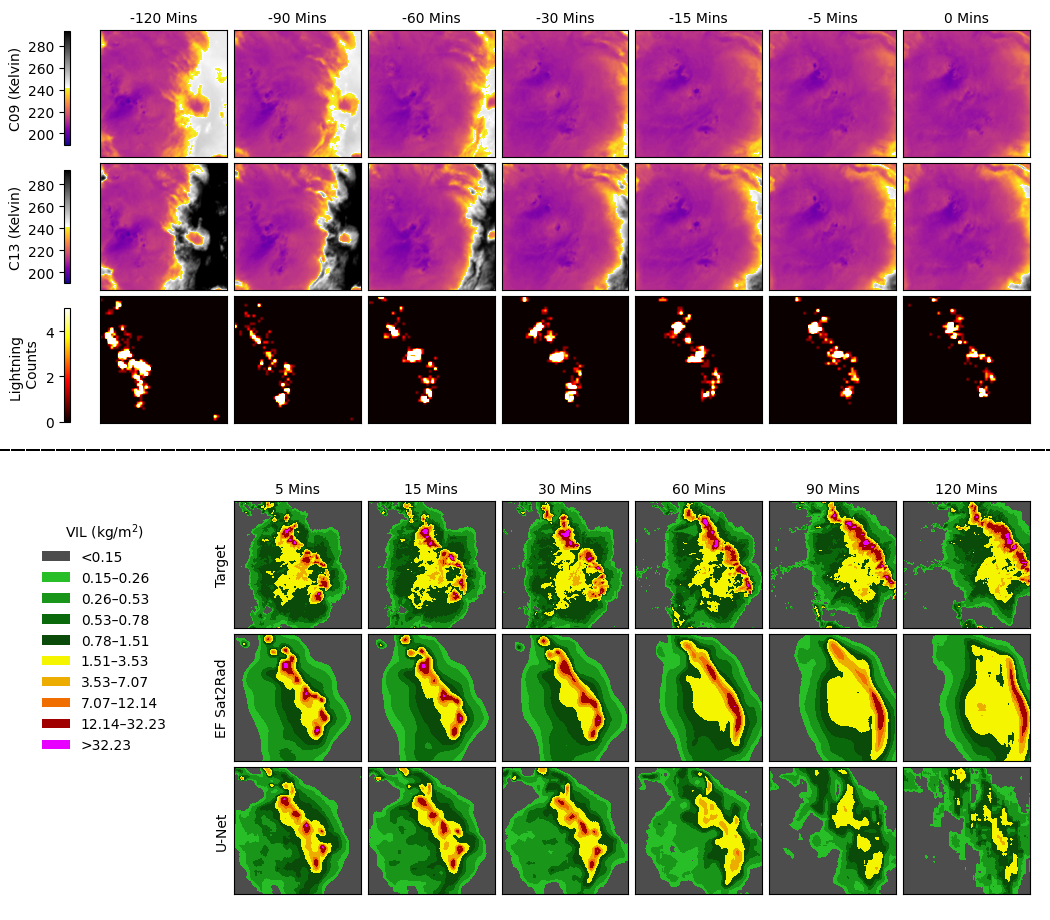}
	\caption{Same as Fig. \ref{fig:modelOut}, but for an event with ID=R19061502217821 in the SEVIR catalog.}	% , plotted with the same colorbar,
	%	\label{fig:modelOut_supp1}
\end{figure*}

\begin{figure*}[h] % Model output
	\centering
	\includegraphics[width=.7\linewidth,keepaspectratio]{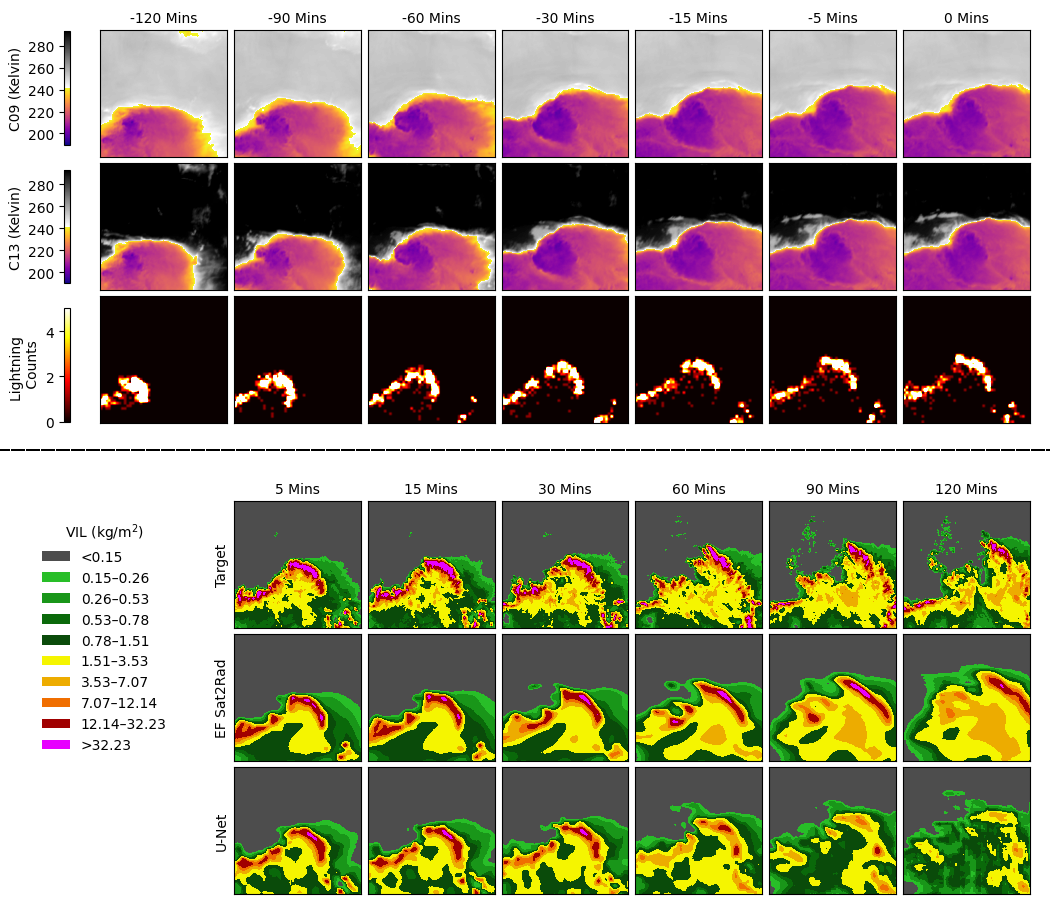}
	\caption{Same as Fig. \ref{fig:modelOut}, but for an event with ID=R19081806217832 in the SEVIR catalog.}	% , plotted with the same colorbar,
	%	\label{fig:modelOut_supp1}
\end{figure*}

\begin{figure*}[h] % Model output
	\centering
	\includegraphics[width=.7\linewidth,keepaspectratio]{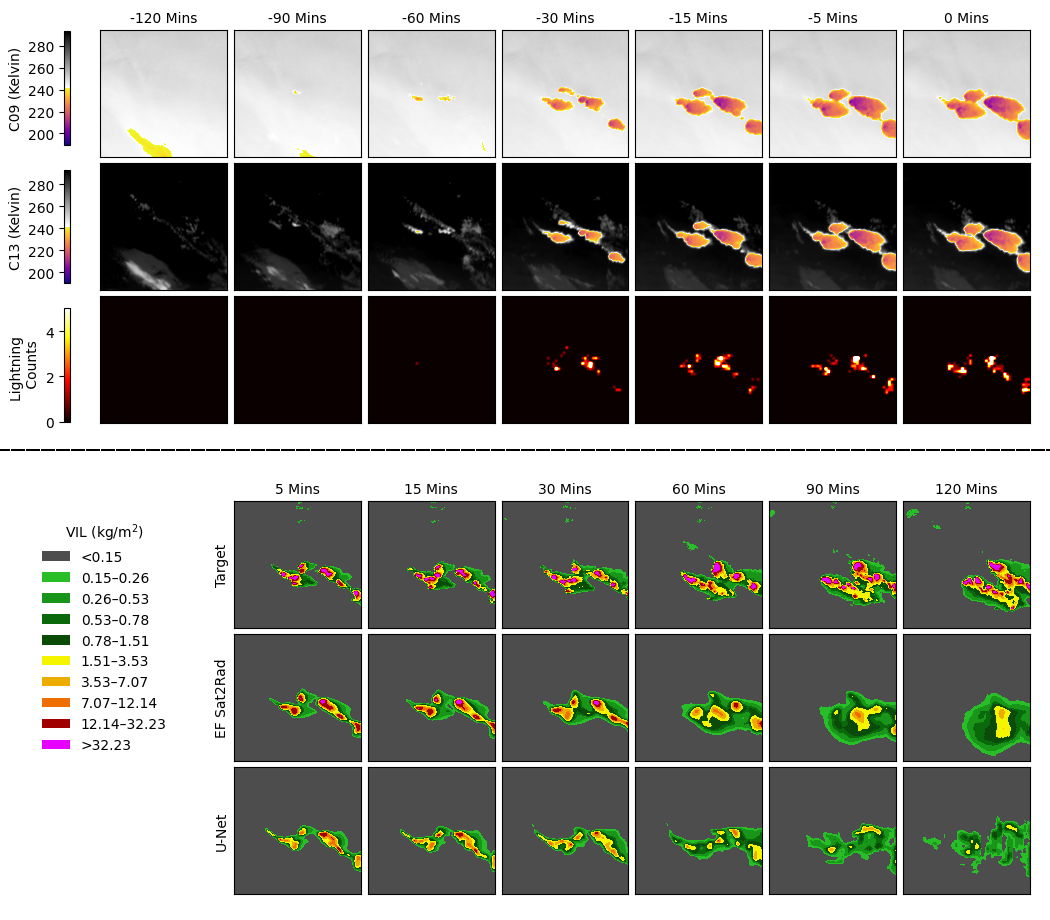}
	\caption{Same as Fig. \ref{fig:modelOut}, but for an event with ID=R19092800327848 in the SEVIR catalog.}	% , plotted with the same colorbar,
	%	\label{fig:modelOut_supp1}
\end{figure*}

\begin{figure*}[h] % Model output
	\centering
	\includegraphics[width=.7\linewidth,keepaspectratio]{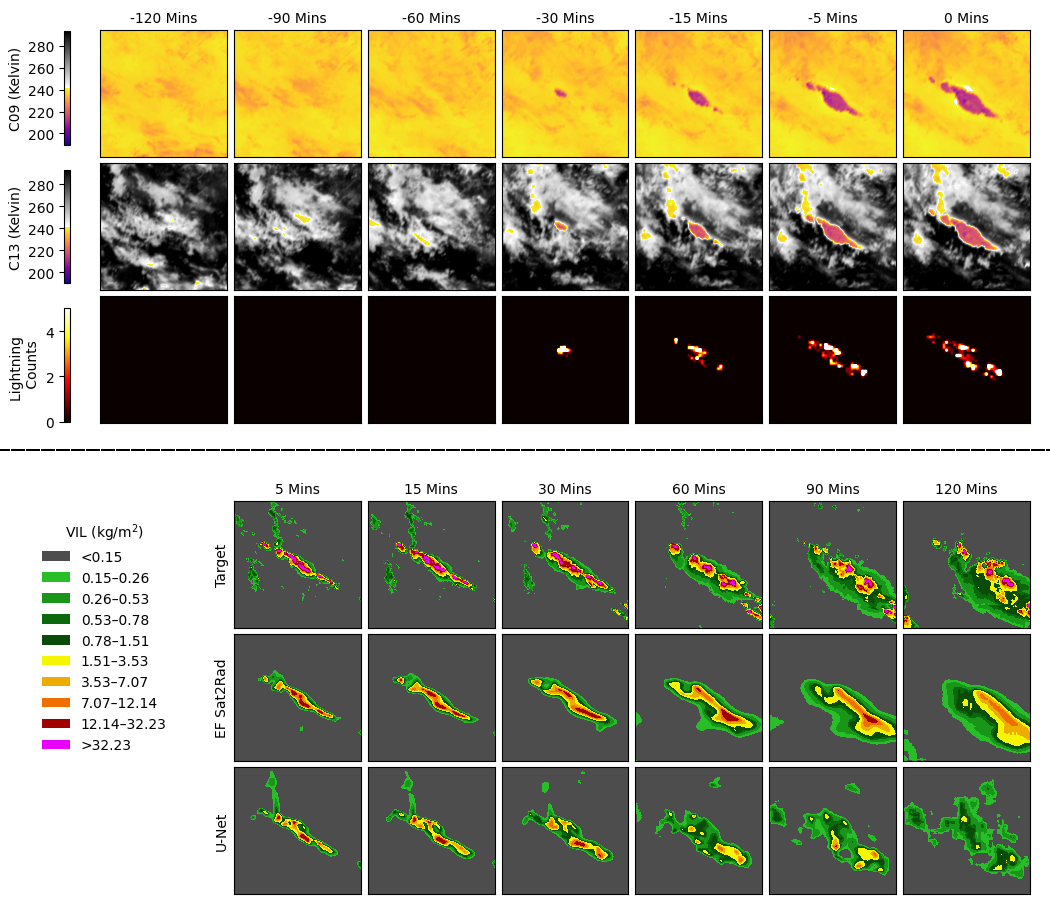}
	\caption{Same as Fig. \ref{fig:modelOut}, but for a tornado event with ID=S831869 in the SEVIR catalog.}	% , plotted with the same colorbar,
	%	\label{fig:modelOut_supp1}
\end{figure*}

\clearpage
\subsection{Worst examples}

%\textbf{Best}

\begin{figure*}[h] % Model output
	\centering
	\includegraphics[width=.7\linewidth,keepaspectratio]{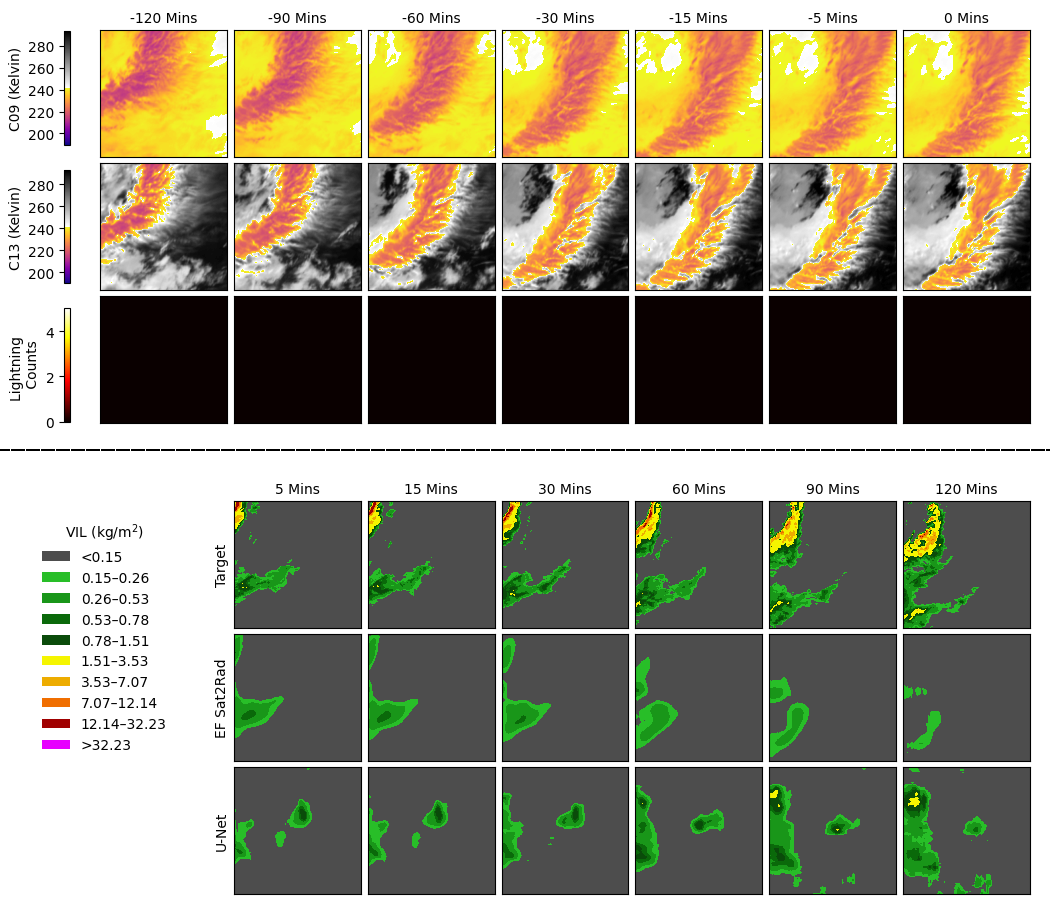}
	\caption{Same as Fig. \ref{fig:modelOut}, but for an event with ID=R19070915538410 in the SEVIR catalog.}	% , plotted with the same colorbar,
	%	\label{fig:modelOut_supp1}
\end{figure*}

\begin{figure*}[h] % Model output
	\centering
	\includegraphics[width=.7\linewidth,keepaspectratio]{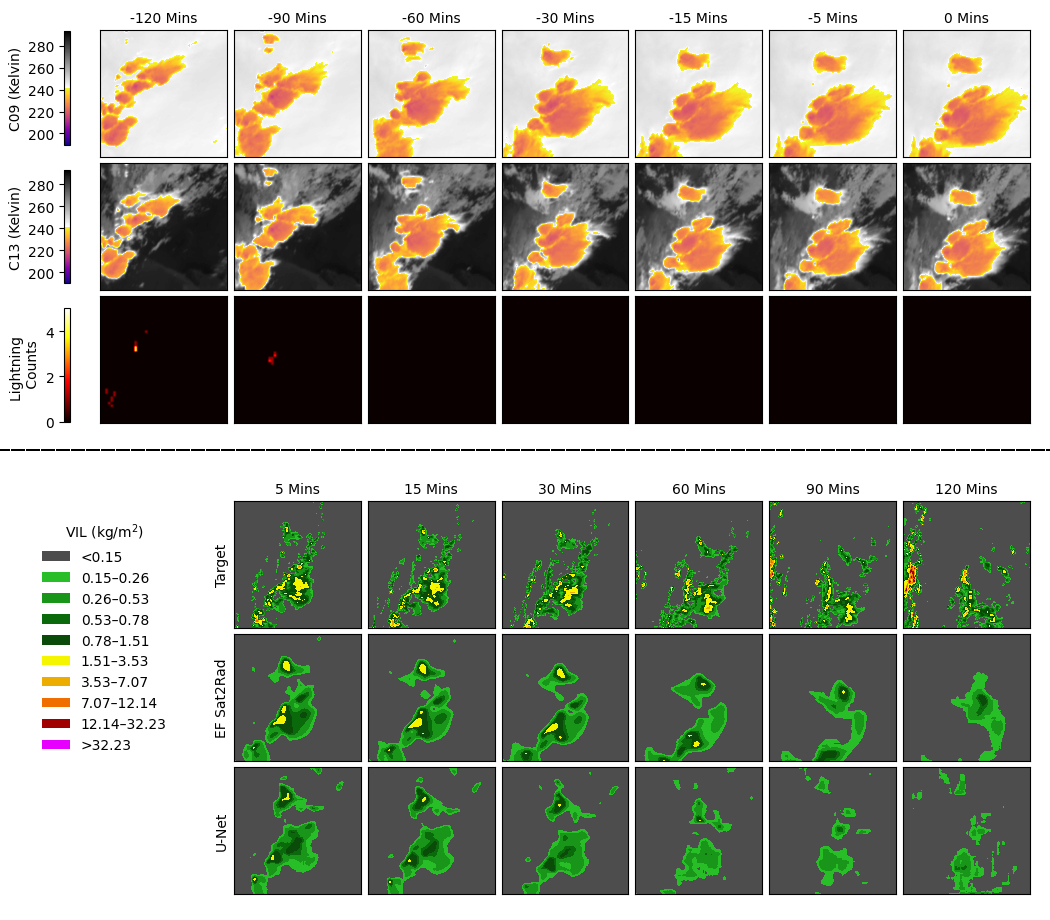}
	\caption{Same as Fig. \ref{fig:modelOut}, but for an event with ID=R19092111378468 in the SEVIR catalog.}	% , plotted with the same colorbar,
	%	\label{fig:modelOut_supp1}
\end{figure*}

\begin{figure*}[h] % Model output
	\centering
	\includegraphics[width=.7\linewidth,keepaspectratio]{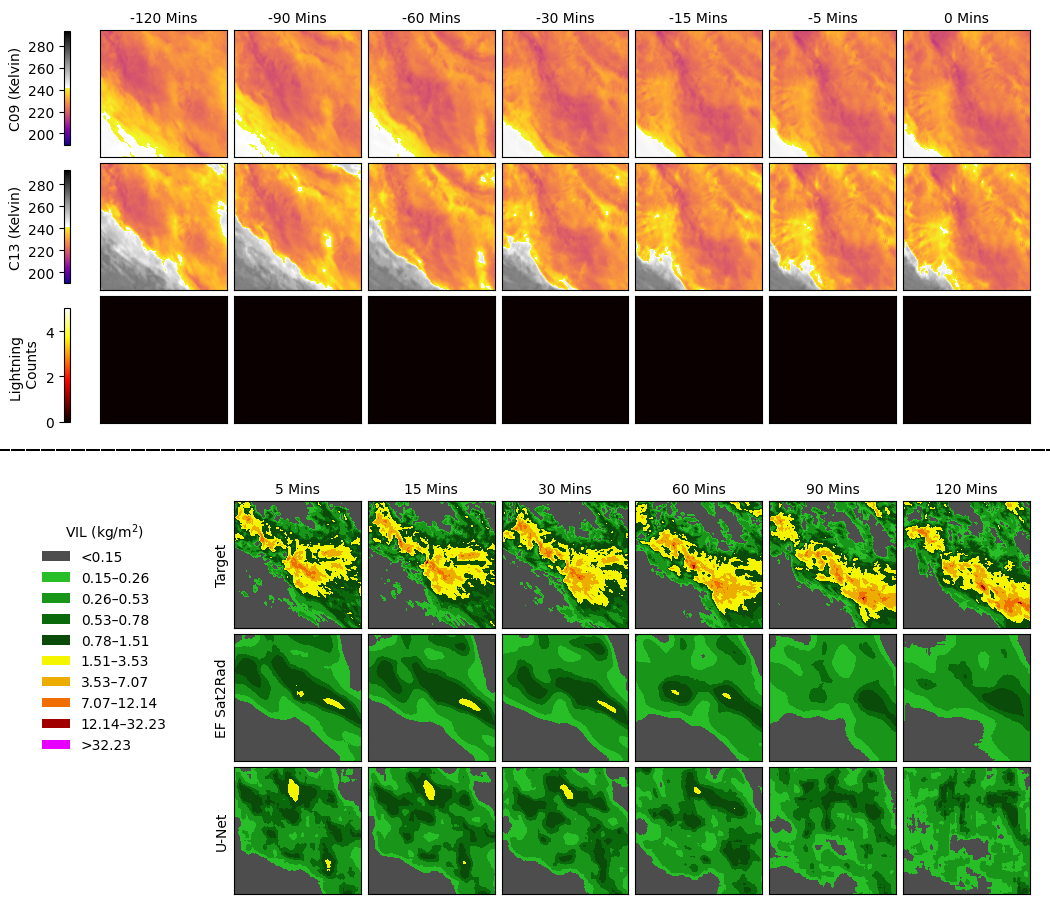}
	\caption{Same as Fig. \ref{fig:modelOut}, but for an event with ID=R19103005428034 in the SEVIR catalog.}	% , plotted with the same colorbar,
	%	\label{fig:modelOut_supp1}
\end{figure*}

\begin{figure*}[h] % Model output
	\centering
	\includegraphics[width=.7\linewidth,keepaspectratio]{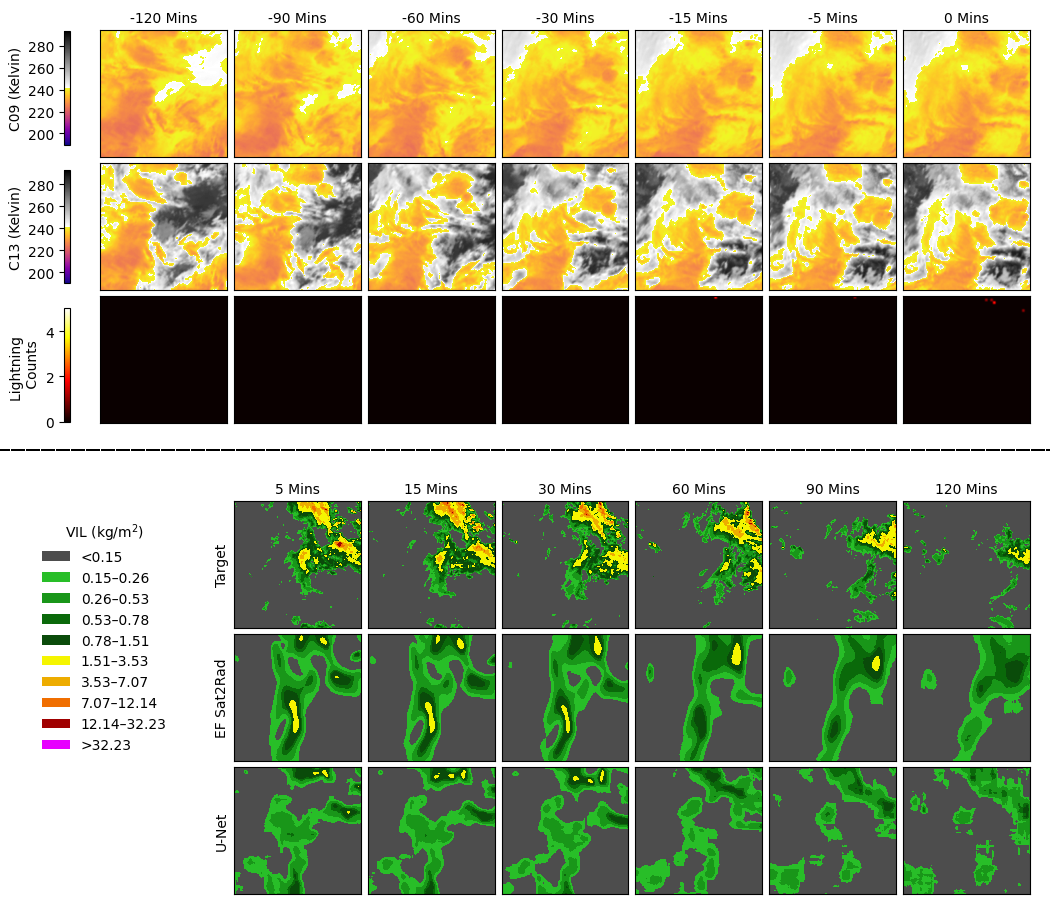}
	\caption{Same as Fig. \ref{fig:modelOut}, but for an event with ID=R19112016107635 in the SEVIR catalog.}	% , plotted with the same colorbar,
	%	\label{fig:modelOut_supp1}
\end{figure*}

\begin{figure*}[h] % Model output
	\centering
	\includegraphics[width=.7\linewidth,keepaspectratio]{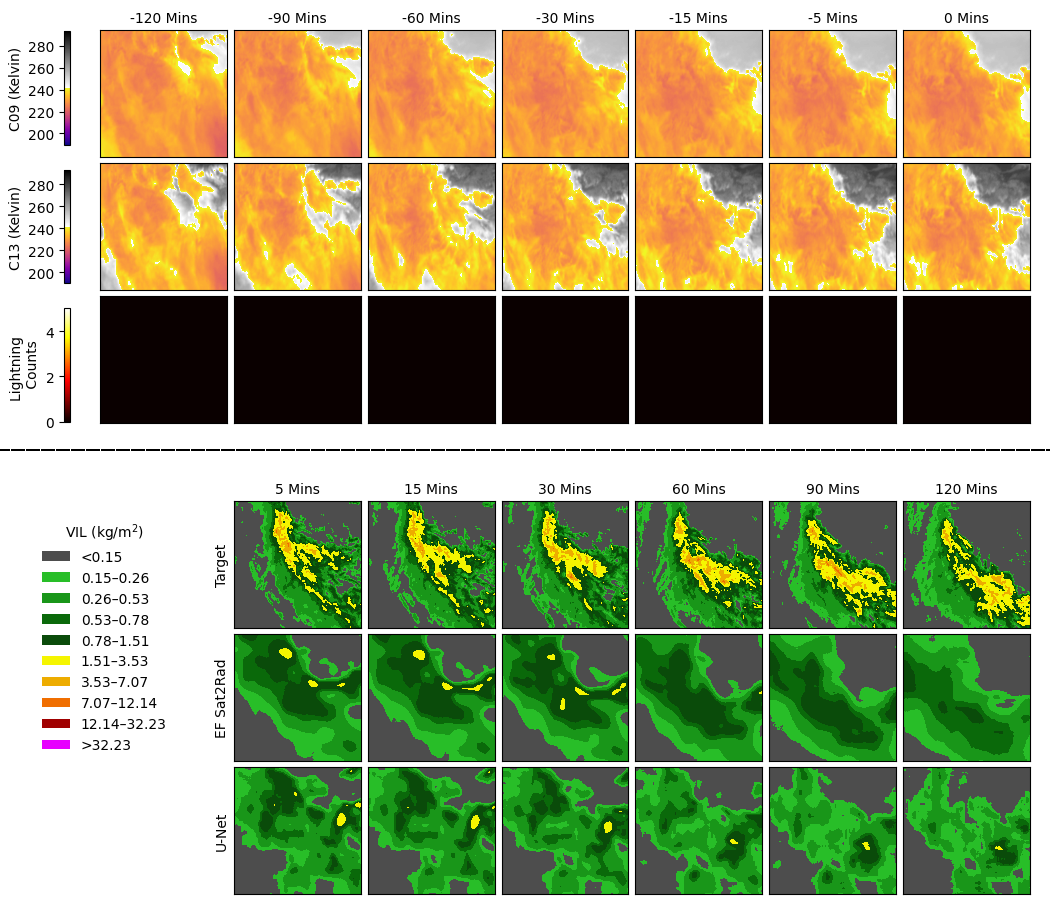}
	\caption{Same as Fig. \ref{fig:modelOut}, but for an event with ID=R19112421488415 in the SEVIR catalog.}	% , plotted with the same colorbar,
	%	\label{fig:modelOut_supp1}
\end{figure*}

\clearpage
\subsection{Random examples}

%\textbf{Best}

\begin{figure*}[h] % Model output
	\centering
	\includegraphics[width=.7\linewidth,keepaspectratio]{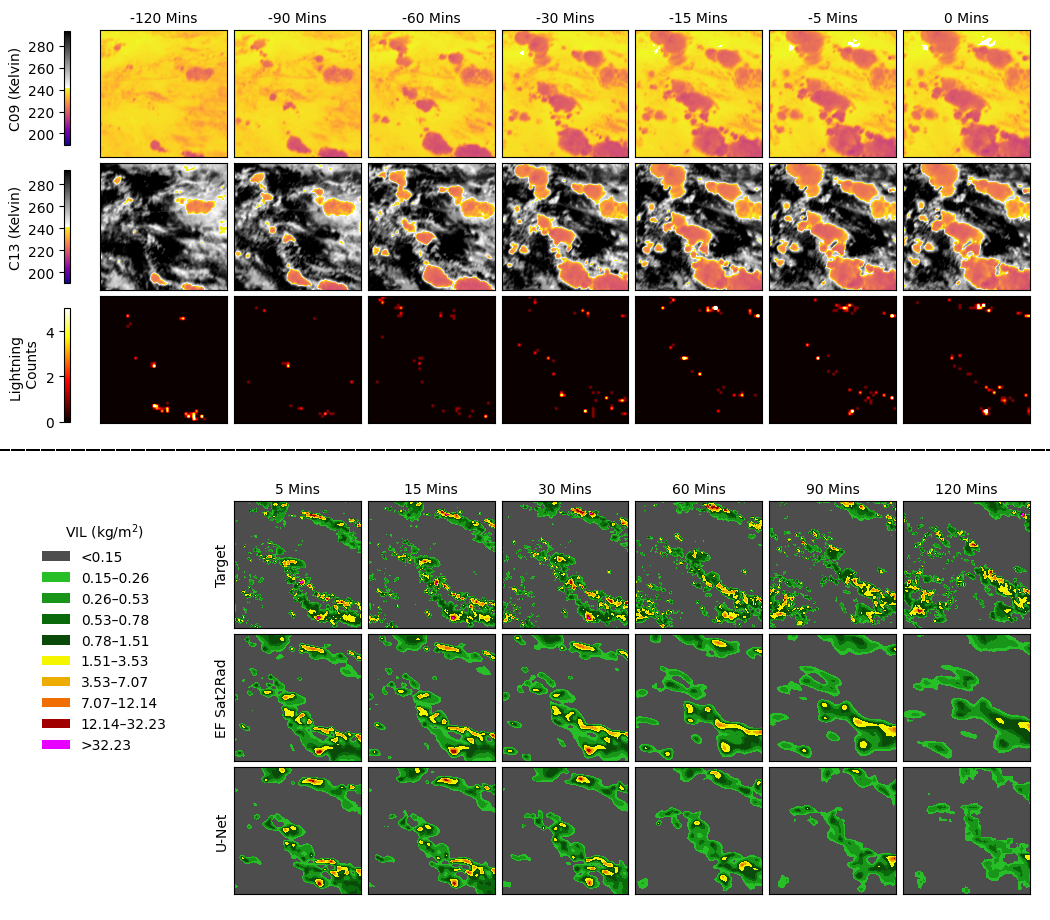}
	\caption{Same as Fig. \ref{fig:modelOut}, but for an event with ID=R19061119368410 in the SEVIR catalog.}	% , plotted with the same colorbar,
	%	\label{fig:modelOut_supp1}
\end{figure*}

\begin{figure*}[h] % Model output
	\centering
	\includegraphics[width=.7\linewidth,keepaspectratio]{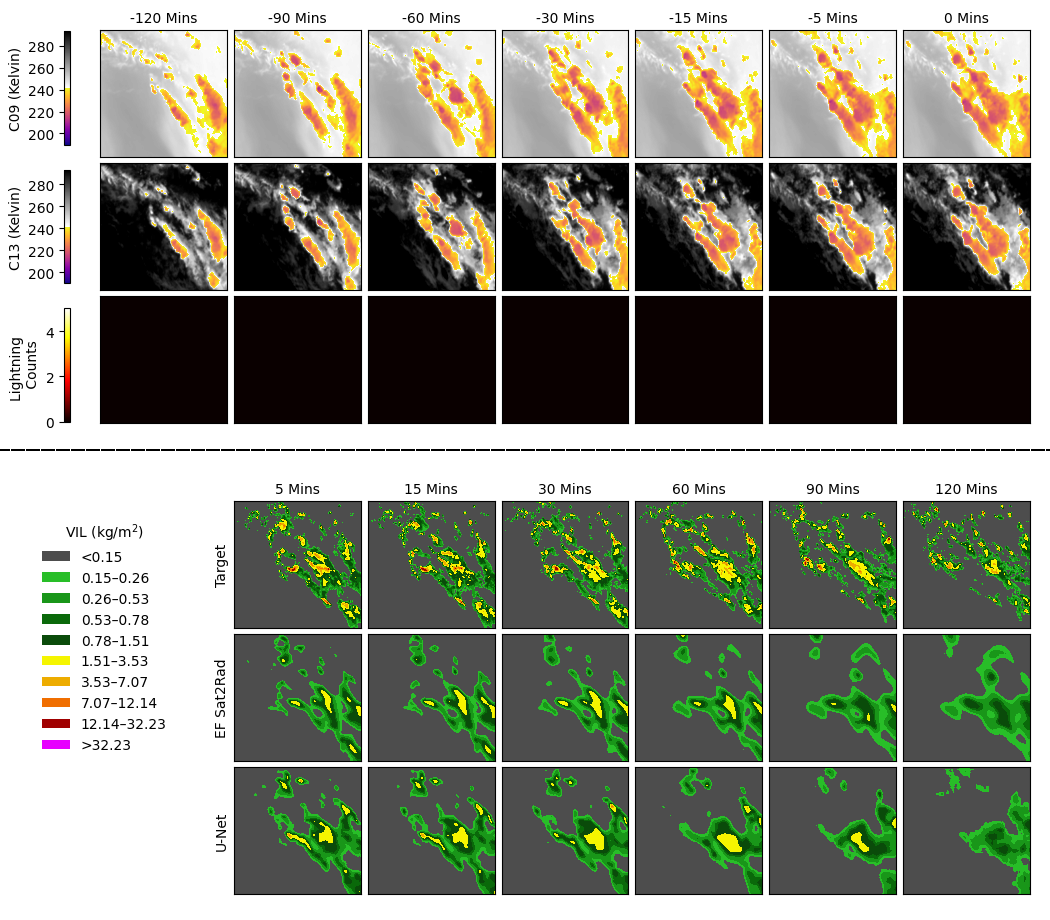}
	\caption{Same as Fig. \ref{fig:modelOut}, but for an event with ID=R19072715377452 in the SEVIR catalog.}	% , plotted with the same colorbar,
	%	\label{fig:modelOut_supp1}
\end{figure*}

\begin{figure*}[h] % Model output
	\centering
	\includegraphics[width=.7\linewidth,keepaspectratio]{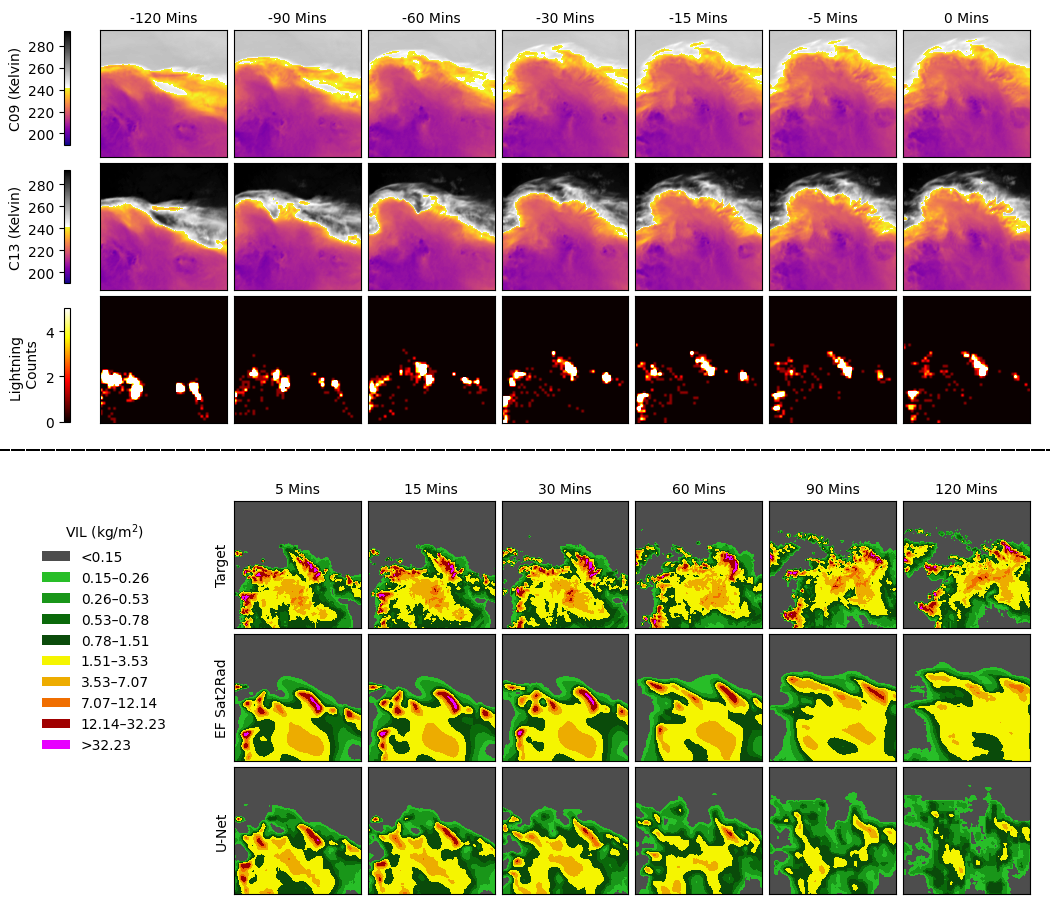}
	\caption{Same as Fig. \ref{fig:modelOut}, but for an event with ID=R19081608527826 in the SEVIR catalog.}	% , plotted with the same colorbar,
	%	\label{fig:modelOut_supp1}
\end{figure*}

\begin{figure*}[h] % Model output
	\centering
	\includegraphics[width=.7\linewidth,keepaspectratio]{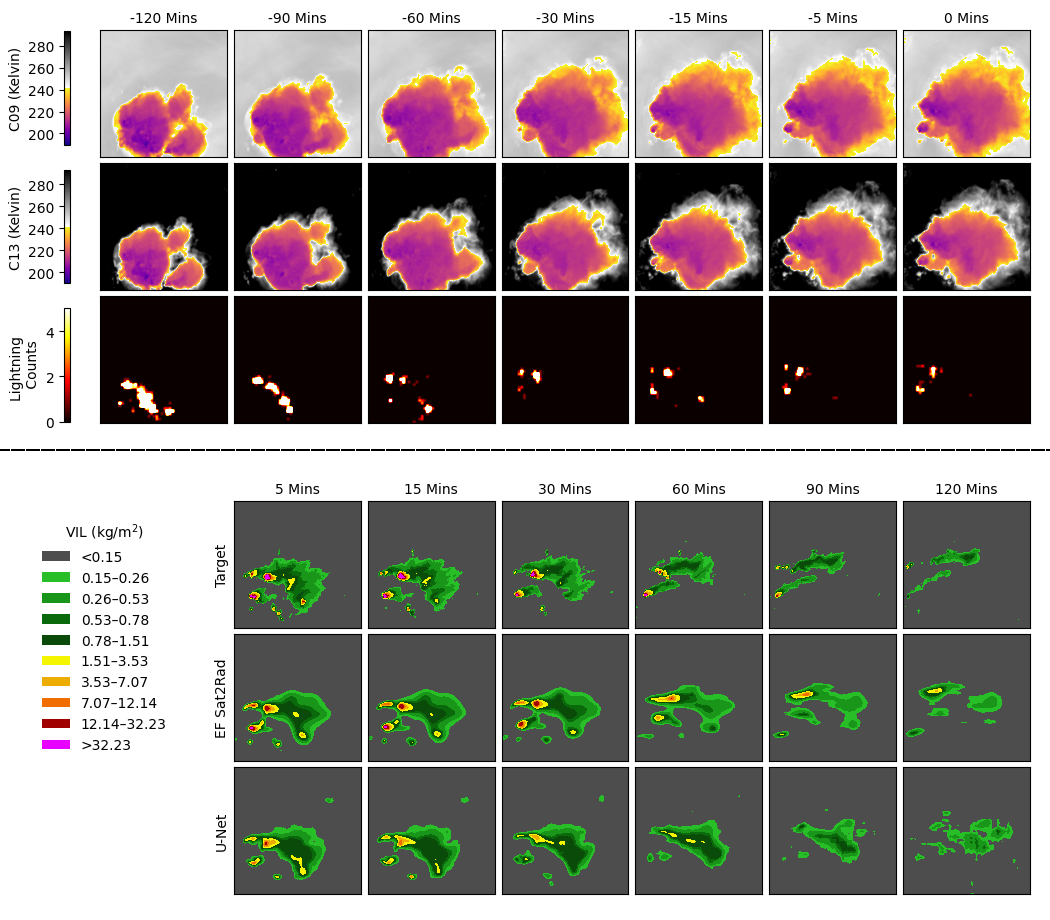}
	\caption{Same as Fig. \ref{fig:modelOut}, but for an event with ID=R19083001247563 in the SEVIR catalog.}	% , plotted with the same colorbar,
	%	\label{fig:modelOut_supp1}
\end{figure*}

\begin{figure*}[h] % Model output
	\centering
	\includegraphics[width=.7\linewidth,keepaspectratio]{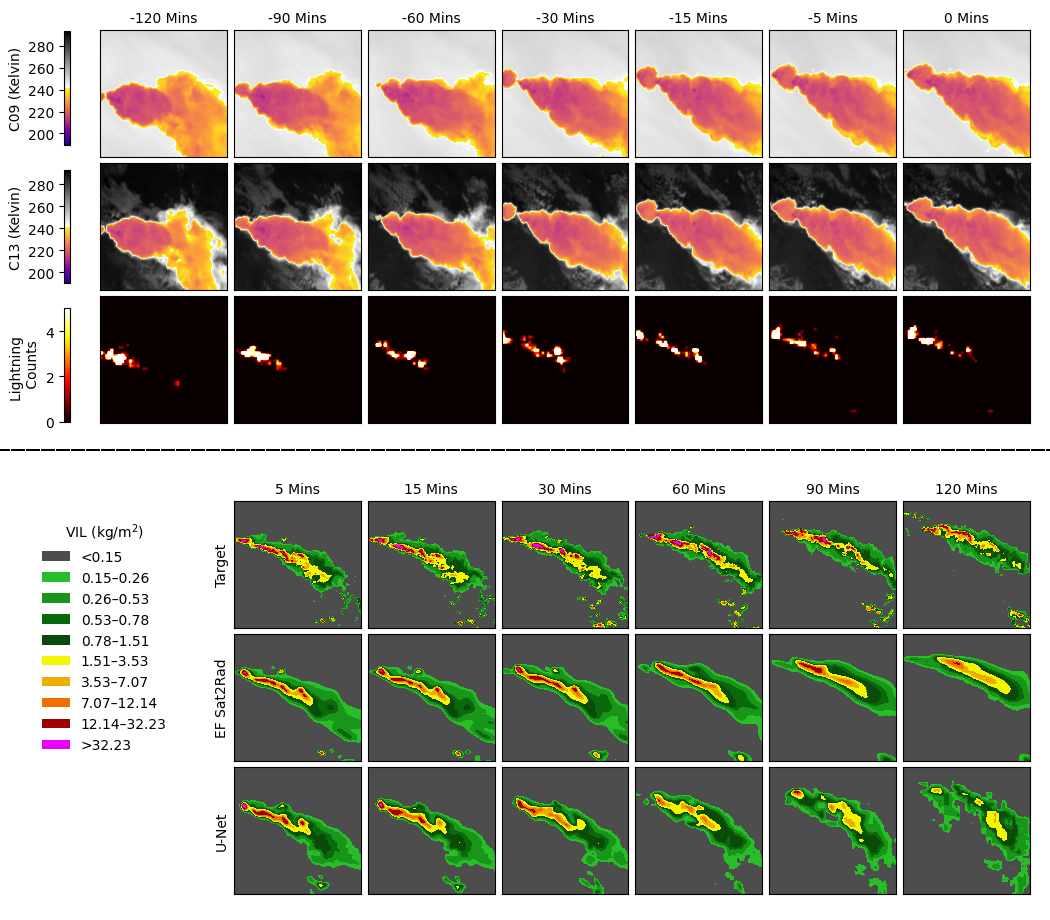}
	\caption{Same as Fig. \ref{fig:modelOut}, but for an event with ID=R19083001248320 in the SEVIR catalog.}	% , plotted with the same colorbar,
	%	\label{fig:modelOut_supp1}
\end{figure*}

\begin{figure*}[h] % Model output
	\centering
	\includegraphics[width=.7\linewidth,keepaspectratio]{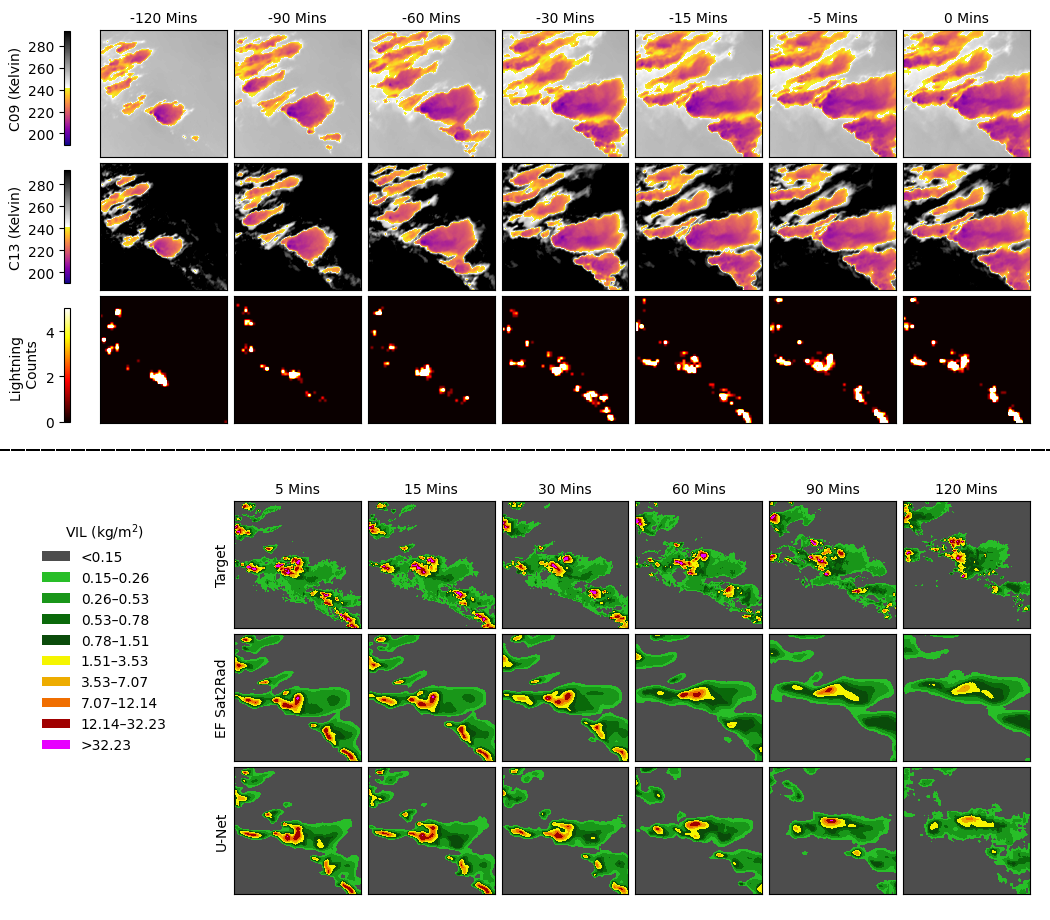}
	\caption{Same as Fig. \ref{fig:modelOut}, but for an event with ID=R19100622287527 in the SEVIR catalog.}	% , plotted with the same colorbar,
	%	\label{fig:modelOut_supp1}
\end{figure*}

\begin{figure*}[h] % Model output
	\centering
	\includegraphics[width=.7\linewidth,keepaspectratio]{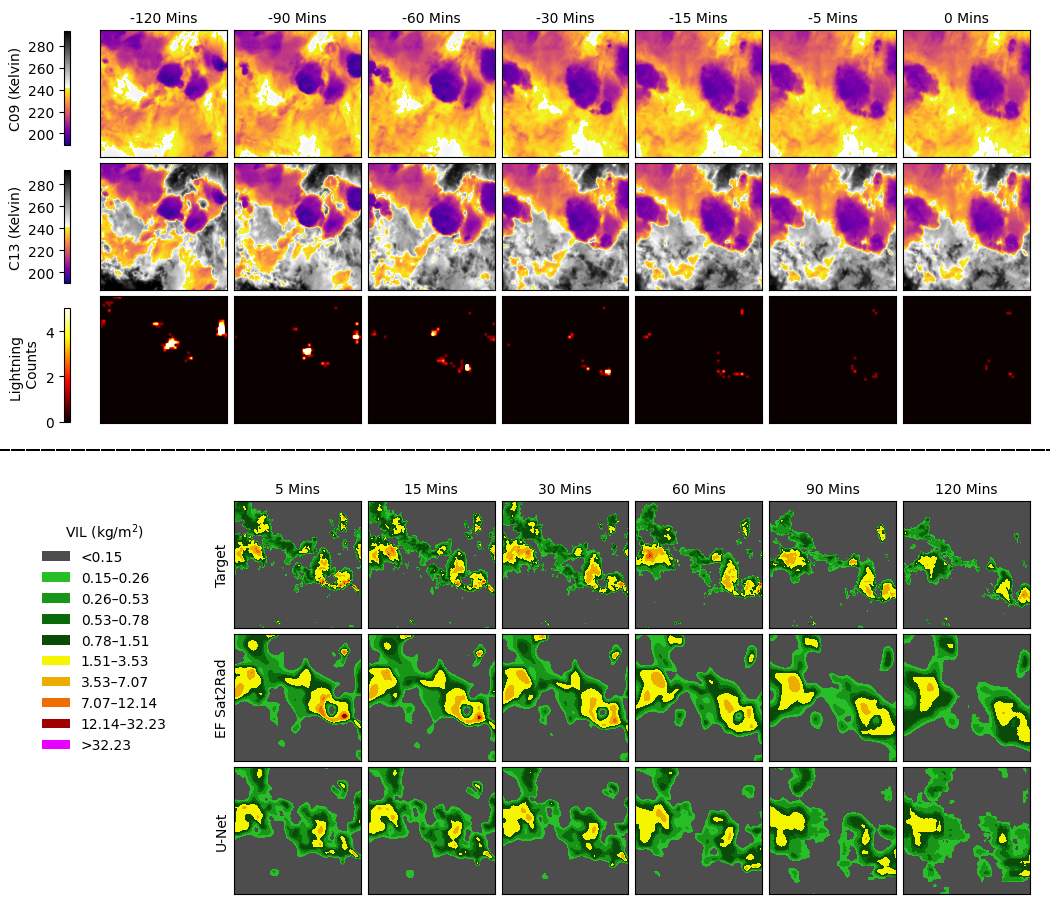}
	\caption{Same as Fig. \ref{fig:modelOut}, but for an event with ID=R19100900257102 in the SEVIR catalog.}	% , plotted with the same colorbar,
	%	\label{fig:modelOut_supp1}
\end{figure*}

\begin{figure*}[h] % Model output
	\centering
	\includegraphics[width=.7\linewidth,keepaspectratio]{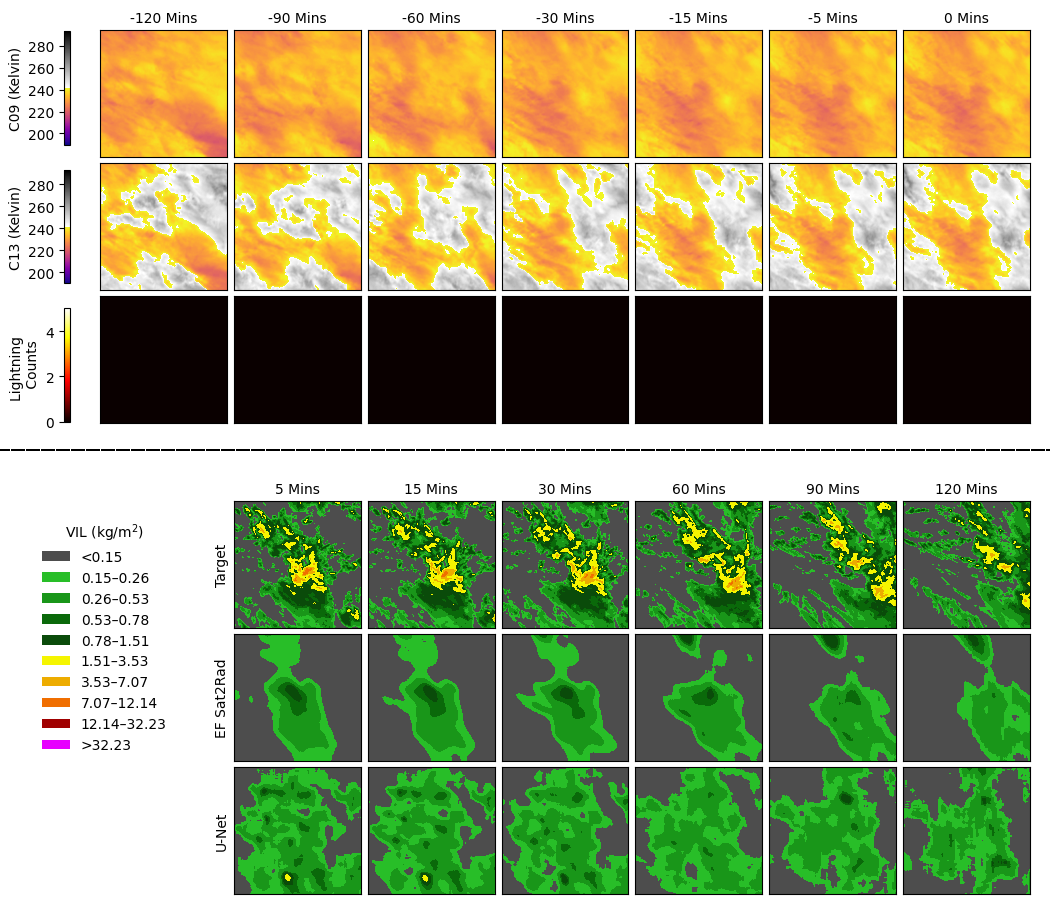}
	\caption{Same as Fig. \ref{fig:modelOut}, but for an event with ID=R19111123427753 in the SEVIR catalog.}	% , plotted with the same colorbar,
	%	\label{fig:modelOut_supp1}
\end{figure*}

\begin{figure*}[h] % Model output
	\centering
	\includegraphics[width=.7\linewidth,keepaspectratio]{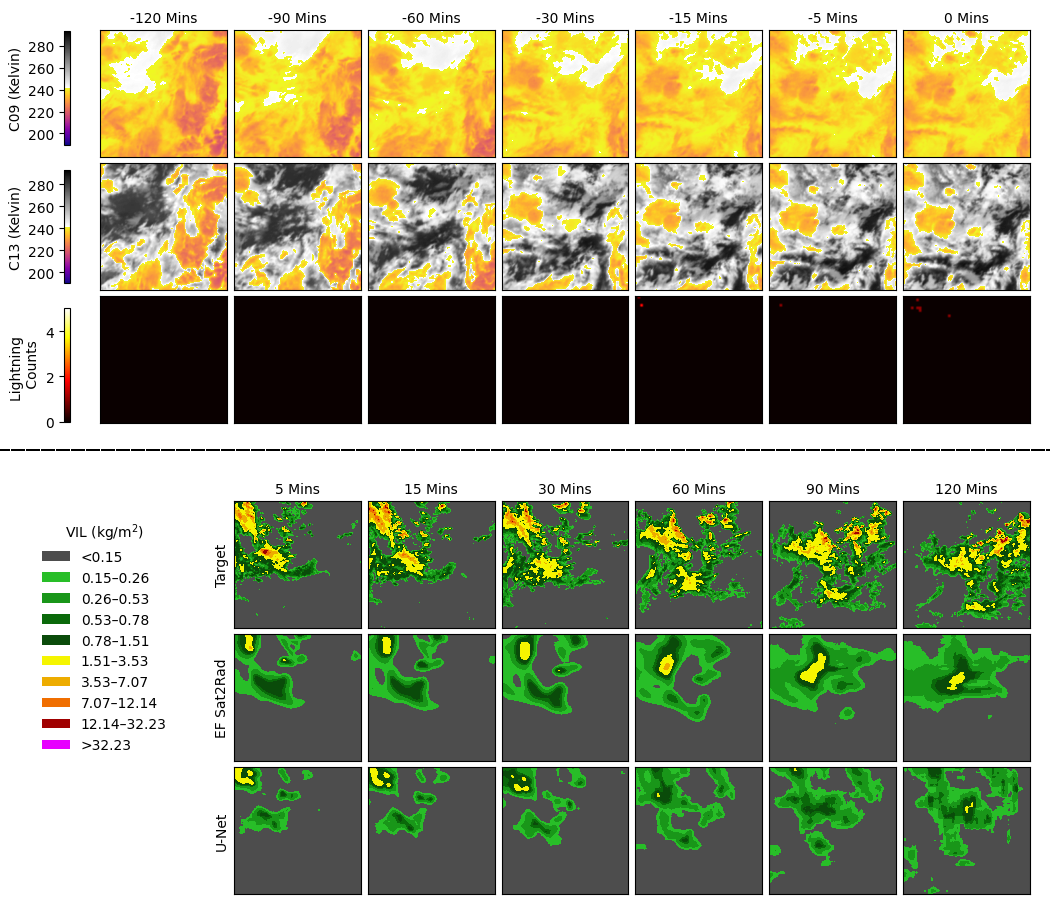}
	\caption{Same as Fig. \ref{fig:modelOut}, but for an event with ID=R19112016107658 in the SEVIR catalog.}	% , plotted with the same colorbar,
	%	\label{fig:modelOut_supp1}
\end{figure*}

\begin{figure*}[h] % Model output
	\centering
	\includegraphics[width=.7\linewidth,keepaspectratio]{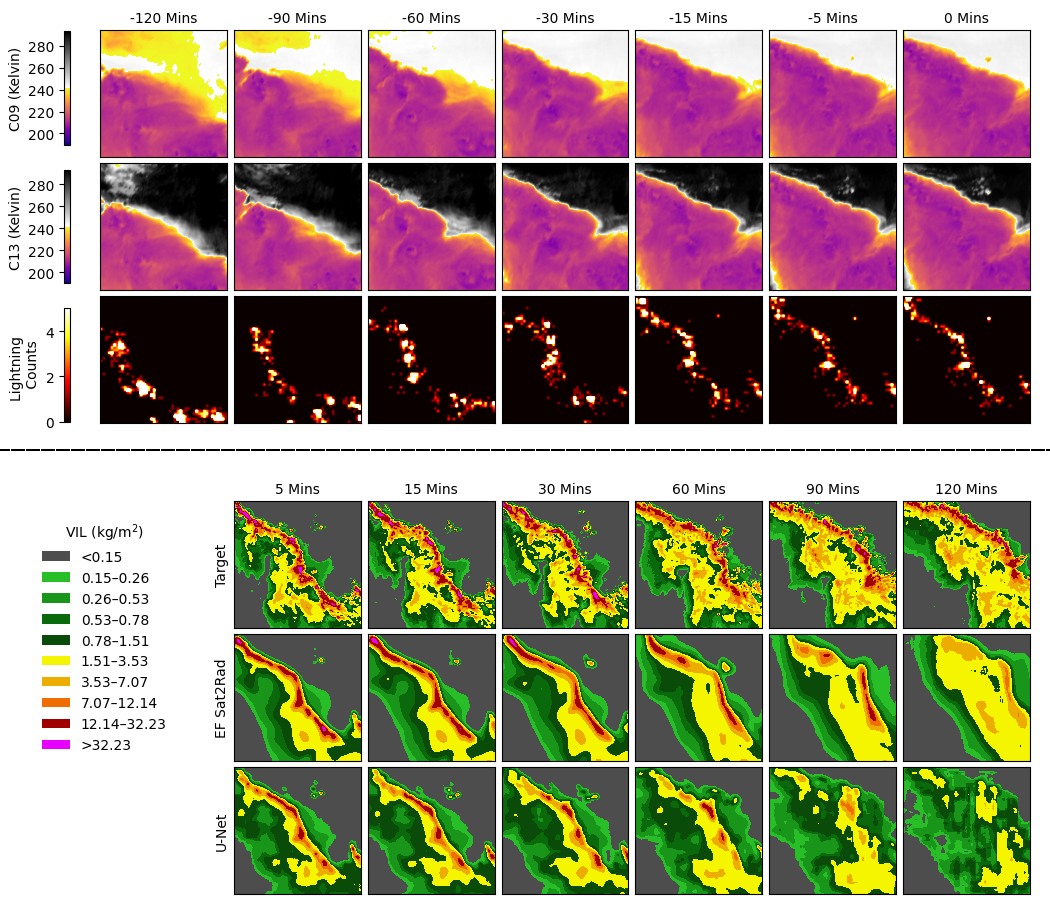}
	\caption{Same as Fig. \ref{fig:modelOut}, but for a thunderstorm wind event with ID=S829040 in the SEVIR catalog.}	% , plotted with the same colorbar,
	%	\label{fig:modelOut_supp1}
\end{figure*}

%TC:endignore
\end{document}